\documentclass[10pt,twocolumn,superscriptaddress,longbibliography,prx]{revtex4-2}

\usepackage{amsmath, bm, amsfonts, amssymb}     
\usepackage{graphicx}                       
\usepackage{xfrac}                          
\usepackage{grffile}                        
\usepackage{color,colortbl}
\usepackage{xcolor}

\usepackage[colorlinks, citecolor=blue, urlcolor=red, linkcolor=blue]{hyperref}

\newcommand{\gammabar}{\bar{\gamma}}
\newcommand{\gdot}[0] {\dot{\gamma}}

\newcommand{\vecv}[1]{\mathbf{{#1}}}
\newcommand{\tens}[1]{\mathbf{{#1}}}

\newcommand{\be}{\begin{equation}}
\newcommand{\ee}{\end{equation}}
\newcommand{\bea}{\begin{eqnarray}}
\newcommand{\eea}{\end{eqnarray}}

\begin{document}

\title{Yielding, shear banding and brittle failure of amorphous materials}
\author{Joseph Pollard and Suzanne M. Fielding}
\affiliation{Department of Physics, Durham University, Science Laboratories,
  South Road, Durham DH1 3LE, UK}

\begin{abstract}

Widespread processes in nature and technology are governed by the
dynamical transition whereby a material in an initially solid-like
state, whether soft or hard, then yields. Major unresolved questions
concern whether any material will yield smoothly and gradually
(``ductile'' behaviour) or fail abruptly and catastrophically
(``brittle'' behaviour); the roles of sample annealing, disorder and
shear band formation in the onset of yielding and failure; and, most
importantly from a practical viewpoint, whether any impending
catastrophic failure can be predicted before it happens. We address
these questions by studying theoretically the yielding of slowly
sheared athermal amorphous materials, within a minimal mesoscopic
lattice elastoplastic description. Our contributions are
fourfold. First, we elucidate whether yielding will be ductile or
brittle, for any given level of sample annealing prior to shear. For
highly annealed samples, we find brittle yielding for all samples
sizes. For poorly annealed samples we uncover an important dependence
on the size of the sample of material being sheared, with ductile
yielding for small samples, and brittle yielding only for large system
sizes. Second, we show that yielding comprises two distinct stages: a
pre-failure stage, in which small levels of strain heterogeneity
slowly accumulate within the material, followed by a catastrophic
brittle failure event, in which a shear band quickly
propagates across the sample via a cooperating line of (individually)
localised plastic events. Third, we provide an exact expression for
the slowly growing level of strain heterogeneity in the pre-failure
stage, expressed in terms of the macroscopically measured
stress-strain curve and the sample size, and in excellent agreement
with our simulation results. Fourth, we elucidate the basic mechanism
via which a shear band then nucleates, in terms of the onset of
cooperativity between plastic events. We furthermore provide an
expression for the probability distribution of shear strains at which
failure occurs, expressed in terms of the sample size and the disorder
inherent in the sample, as determined by the degree of annealing prior
to shear. Importantly, this indicates a possible route to predicting
impending sudden material failure, before it occurs.

\end{abstract}

\maketitle

\section{Introduction}
\label{sec:intro}

Amorphous materials~\cite{ISI:000288917400017,nicolas2018deformation}
include yield stress
fluids~\cite{ISI:000407999000001,ISI:000266878700032} and soft glassy
materials~\cite{ISI:A1997WM06400048} such as dense colloids,
emulsions, foams and microgels, as well as hard materials such as
molecular and metallic
glasses~\cite{ISI:000374617600037,ISI:000320906700001}.  When
experiencing low loads or small deformations, such materials typically
behave in a solid-like way. At higher loads or larger deformations,
they then yield plastically. Indeed, numerous processes in nature and
technology are governed by the dynamical transition whereby a material in
an initially solid-like state yields, whether suddenly or
gradually. Examples include the restarting of a pipeline of waxy crude
oil~\cite{norrman2016axial}; the rising of bubbles in radioactive
sludge~\cite{gens2009full}; the spreading of fresh
cement~\cite{chidiac2009plastic,banfill1981viscometric}; the yielding
of food during chewing~\cite{fischer2011rheology}; the catastrophic
material failure of metallic glasses~\cite{ISI:000320906700001};
geological processes such as landslides and lava
flows~\cite{ISI:000175054700034,mader2013rheology}; and the reshaping
of biological tissue under internal stresses caused by active
processes such as cell division or cell
death~\cite{ranft2010fluidization,gonzalez2012soft,park2015unjamming}.

In the context of yield stress fluids and soft glassy materials, a
common rheological protocol consists of subjecting a sample at some
time $t=0$ to the switch-on of a shear of some strain rate $\gdot$,
which is held constant thereafter. A typical shear stress response
$\Sigma(\gamma=\gdot t)$ then shows an initially elastic solid-like
regime in which the stress increases linearly with strain $\gamma$, followed by
a stress maximum. This signifies the onset of yielding, and the stress
then declines towards a constant in the final steadily flowing state.
The flow field across the sample often becomes highly heterogeneous as
yielding sets
in~\cite{ISI:000295085700080,ISI:000277945900061,ISI:000301801100015,ISI:000332461800012,ISI:000268689400009,ISI:000280140800011,ISI:000261891200077,ISI:000365222200015}:
shear bands form, with layer normals in the flow-gradient
direction. These bands can take hours finally to heal away, leaving a
homogeneously fluidised ultimate flowing state. Exploiting recently
developed techniques for tracking the motions inside a
fluid~\cite{ISI:000254405700004,ISI:000263613000001,ISI:000321273500022},
experiments have mapped the complicated spatio-temporal processes
involved~\cite{ISI:000295085700080,ISI:000277945900061,ISI:000301801100015,ISI:000332461800012,ISI:000268689400009,ISI:000280140800011,ISI:000261891200077,ISI:000365222200015,ISI:000357577400001,ISI:000379585600007,ISI:000374962100013,ISI:000320314800001,ISI:000229261800006,ISI:000317915200028,ISI:000300970700003,ISI:000287095800012,ISI:000254473800079}. Besides
the gradual yielding and fluidisation just described, some soft
materials instead fail suddenly, via brittle
frature~\cite{ISI:000381495100025,ISI:A1996TR32200033,ISI:000073082400044,ISI:000334096300032,ISI:000265285700042,ISI:000341260100009}.

Alongside the complex fluids and soft solids just described, some
notably similar phenomenology is seen in harder materials such as
polymeric and metallic
glasses~\cite{ISI:000374617600037,chen2008mechanical,boyce1988large,anand2012large,ward2012mechanical}. These
likewise show initially solid-like behaviour at low strains, followed
by yielding with a strong associated tendency for strain localisation
and shear banding~\cite{ISI:000320906700001}. But whereas yield stress
fluids can then flow indefinitely post-yield without losing their
ability to return to a solid state when later unloaded, in metallic
glasses the formation of shear bands typically leads to brittle
yielding and catastrophic sample failure, thereby limiting the
material's strength. Different scenarios for its onset have been
discussed~\cite{ISI:000320906700001,cheng2011intrinsic,tian2012approaching}. These
include homogeneous nucleation~\cite{shi2006atomic}, in which local
plastic events triggered within the material under shear spatially
cooperate to form a system-spanning shear band, once a percolation
threshold for their link-up has been achieved as a function of
increasing stress. The stress needed for this can be rather large,
however, so in practice shear bands often nucleate heterogeneously at
stress concentrations, for example around any indentations on the
sample's surface~\cite{shi2007stress,csopu2017atomic,vasoya2016notch}. The important
role of sample size in modifying shear banding formation and brittle
yielding has also been studied~\cite{yang2012size,wang2016sample}.  In
particular, the yield strength of metallic glasses has been found to
increase with decreasing sample size: a phenomenon suggested to arise
from the reduced number of sites for shear band nucleation as system
size decreases.

Theoretical studies of yielding in amorphous materials have ranged
from microscopic through mesoscopic and up to the continuum level. In
the context of continuum modelling of the rheology of complex fluids,
it can be predicted within a minimal set of generalised constitutive
assumptions that a state of initially homogeneous shear must
generically become unstable to the formation of shear bands once the
maximum of stress as a function of strain is attained, and the
material first starts to
flow~\cite{ISI:000384392300002,ISI:000329357400005,ISI:000315141600016,ISI:000316969500006,ISI:000293534200007,ISI:000286879900011,ISI:000344142000001}. For
samples subject instead to a constant applied shear stress, shear
bands are likewise generically predicted to form as the initial creep
regime terminates and the sample yields. These predictions have been
confirmed
experimentally~\cite{ISI:000295085700080,ISI:000277945900061,ISI:000301801100015,ISI:000332461800012,ISI:000268689400009,ISI:000280140800011,ISI:000261891200077,ISI:000365222200015},
by molecular
simulations~\cite{ISI:000342206200002,ISI:000386386400004,ISI:000384392300003,ISI:000246210200042,varnik-jcp-120-2788-2004,vasisht2020computational,vasisht2020emergence}
and in several specific constitutive models of complex fluids
rheology~\cite{ISI:000344142000001,ISI:000322544200022,ISI:000286879900011,ISI:000250675300003,ISI:000278158800014,ISI:000285583500029,ISI:000362909100023,ISI:000296658800003,ISI:000251326200013,ISI:000262976900016}.

At the level of microscopic and mesoscopic modelling, much attention
has been devoted to understanding
the dynamical properties of the finally flowing shear state, in which
the stress has attained a (statistically) steady state as a function
of applied strain. (Any such state may or may not be attainable in
practice for a given sample, being sometimes pre-empted by
catastrophic sample failure.) A unifying picture has emerged in which local plastic events triggered within the material by the
applied shear cooperate to form avalanches that flicker intermittently
across the system, with power-law statistics.  Their dependence on
strain rate, system size and temperature has been carefully
characterised~\cite{ISI:A1997WK49200067,ISI:000370030200003,ISI:000337350100004,ISI:000239483400012,ISI:000342633900033,ISI:000332617600014,ISI:000312846200010,ISI:000244592100027,ISI:000249785900011,ISI:000239425700020,ISI:000308394900010,ISI:000328697400002}.

Beyond this steadily flowing state, attention is increasingly turning to understanding the initial onset
of yielding after a shear is first switched on. Several approaches
have been put forward, studying dynamical yielding within a replica
field
theory~\cite{ISI:000370815100008,ISI:000402296700034,ISI:000410885300004};
as a critical point in an elastoplastic
model~\cite{ISI:000362909100023,liu2018creep}; as a directed
percolation transition~\cite{ISI:000386386400004,ISI:000384392300003};
within a random first order transition theory for the glass
transition~\cite{ISI:000309611400031}; as a Gardner transition
~\cite{rainone2015following}; as a spinodal
point~\cite{urbani2017shear}; and within particle simulations that
seed initial weak spots~\cite{ozawa2021rare}. Yielding in oscillatory
shear has been studied using particle
simulations~\cite{ISI:000366295400003,ISI:000382177400007,parmar2019strain},
in elastoplastic models~\cite{liu2020oscillatory} and in energy
landscape models~\cite{sastry2020models}. Yielding following creep
under an applied load was studied using particle simulations in
Ref.~\cite{cabriolu2019precursors}.  Microscopic precursors to
yielding have recently been observed experimentally in soft
materials~\cite{ISI:000429012500051,ISI:000392096800036,ISI:000341025700007}.

With this backdrop, it is clear that yielding represents a problem of
central importance to several areas of physics, with the observation
of common phenomenologies across multiple classes of material
stimulating a search for universal explanations. In the twin contexts
of soft matter physics and fluid dynamics, yielding is crucial to the
question of how complex fluids flow. In statistical physics, it
represents a non-equilibrium phase transition that is only just
starting to be understood from first principles. In materials physics,
it is core to understanding a material's ultimate strength in
principle, and to improving material performance in practice. In
active matter, its role in the reshaping of biological tissues remains
to be elucidated.

A question of key significance to the practical performance of a material is whether it will yield smoothly and gradually
(``ductile'' behaviour) or instead fail abruptly and catastrophically
(``brittle''
behaviour)~\cite{ISI:000436245000061,popovic2018elastoplastic,barlow2020ductile}. The
influence of annealing and disorder in the sample prior to shear, and
its role in determining the formation of shear bands during yielding
and failure, is increasingly being
appreciated~\cite{leishangthem2017yielding,ISI:000231503600036,ISI:000436245000061,popovic2018elastoplastic,barlow2020ductile,vasisht2020emergence}. And
from a practical viewpoint, perhaps the most important question is
whether any impending sudden catastrophic failure can be anticipated,
before it actually occurs, in terms of an identifiable material
property. In this work, we address these questions within a lattice
elastoplastic model~\cite{nicolas2018deformation} that contains only
minimal assumptions, and should therefore capture the mechanics of
quasistatically sheared athermal amorphous materials in a universally
generic way.

Our first contribution will be to carefully elucidate whether yielding
is ductile or brittle, for any given level of sample annealing prior
to shear. For highly annealed samples we find brittle yielding for all
system sizes. In contrast, for poorly annealed samples we demonstrate
an important dependence of the nature of yielding on the size of the
sample being sheared, with ductile yielding for small samples and
brittle yielding only for sizes larger than have previously been
studied theoretically. We thereby resolve an apparent contradiction
between the recent work of
Refs.~\cite{ISI:000436245000061,popovic2018elastoplastic}, on the one
hand, and that of Ref.~\cite{barlow2020ductile} on the other.

Indeed, recent studies of mean field elastoplastic models of
quasistatically sheared athermal amorphous
materials~\cite{ISI:000436245000061,popovic2018elastoplastic} have
suggested ductile and brittle yielding to be separated by a ``random
critical point''. In this scenario, the underlying stress-strain
relation, $\Sigma(\gamma)$, is posited to display a qualitative change
in form as the degree to which the sample has been annealed prior to
shear increases: for poorly annealed samples, $\Sigma(\gamma)$ has an
overshoot but not an overhang, and yielding proceeds in a smoothly
ductile way, whereas for highly annealed samples, it additionally
develops an overhang, leading to catastrophic brittle yielding, with
the stress dropping precipitously once this overhang is reached.
Ductile and brittle yielding are thereby separated by a critical
point, at which the stress-strain curve switches between these two
qualitatively different forms with increasing annealing. Particle
simulations and studies of two-dimensional (2D) lattice elastoplastic
models beyond mean field have been argued to be consistent with this
scenario~\cite{ISI:000436245000061}.

In contrast, calculations performed beyond mean field within an
athermal 1D elastoplastic model~\cite{barlow2020ductile} reported a
discontinuous stress drop for any level of sample annealing, however
small, provided a stress drop indeed exists. They thereby predicted that
yielding will always be brittle in slowly sheared athermal amorphous
materials, contradicting the critical point scenario of
Refs.~\cite{ISI:000436245000061,popovic2018elastoplastic}.

Our work demonstrates that the scenario of
Ref.~\cite{barlow2020ductile}, explored in that work in a simplified
1D elastoplastic model, also obtains in the limit of infinite system
size in 2D lattice elastoplastic models comprising $N\times N$
elastoplastic elements. We achieve this by simulating a range of sample
sizes that has not (to this author's knowledge) been attained in any
previous study, from $N=64$ to $N=4096$. We thereby suggest that
evidence for the random critical point scenario of
Refs.~\cite{ISI:000436245000061,popovic2018elastoplastic}, beyond mean
field, may be a finite size effect, potentially amenable to further
finite size analysis along the lines of that in
Ref.~\cite{singh2020brittle}.

Our next contribution will be to show that yielding comprises two
distinct stages: a pre-failure stage, in which small levels of
strain heterogeneity slowly accumulate across the sample, followed (in
large or highly annealed samples) by a catastrophic brittle failure
event, in which a macroscopic shear band suddenly
propagates across the sample. 

In the pre-failure regime, we provide an exact analytical expression
for the slowly growing level of strain heterogeneity across the
sample, expressed in terms of the macroscopically measurable
stress-strain curve and the sample size. We show this to be in
excellent agreement with our simulation results. We also perform the
counterpart analysis for a mean field elastoplastic model, again
obtaining excellent agreement between our analytical predictions and
simulation results.

Finally, we shall elucidate the mechanism of catastrophic material
failure, in which a  shear band nucleates and spreads quickly
across the sample. Our basic observation is that a single element
first yields plastically and elastically propagates its stress to
other elements via the Eshelby propagator, creating further yielding
events in turn, which cooperatively percolate along a
line~\cite{hieronymus2017shear}. In simulating a 2D lattice with
periodic boundary conditions, our findings necessarily pertain to
homogeneous nucleation. Although in practice failure often occurs
via heterogeneous nucleation of shear bands at  stress
concentrations around surface imperfections, elucidating the basic
physical mechanisms of homogeneous nucleation is an essential
pre-requisite to progress in also understanding heterogeneous
nucleation, and to determining the ultimate theoretical limit of
material strength in the absence of surface imperfections.

We provide an expression for the probability distribution of shear
strains at which a shear band appears and material failure occurs, in terms
of the disorder inherent in the sample, as determined by the degree of
annealing prior to shear. This expression shows good agreement with
our simulation results for highly annealed samples, with progressively
less good agreement for less well annealed samples. Importantly, this
analytical result for the strain at which catastrophic failure occurs
depends only on a single parameter characterising the level of initial
sample annealing, together with the system size. Our findings thereby
provide a possible route to predicting the brittle failure of highly
annealed amorphous materials, before it actually occurs.  This is an
important prediction that we hope will stimulate further work to
validate or disprove it, both experimentally and in particle
simulations of soft and hard amorphous materials.

For highly annealed samples, we demonstrate a weak progression towards
increasing sample strength with decreasing sample size, consistent
with the trend observed in metallic glasses~\cite{wang2016sample}. The
finding of shear band nucleation in our 2D elastoplastic model is
consistent with the insight of Wyart and co-workers in
Ref.~\cite{popovic2018elastoplastic}, who outlined arguments for 
band nucleation in an elastoplastic model based on classical ideas of
fracture mechanics~\cite{anderson2017fracture}.

The paper is structured as follows. In Sec.~\ref{sec:model} we
introduce two elastoplastic models to be studied: a 2D lattice model
with force balance, and its mean field counterpart. These models have been widely studied in the existing literature. In
Sec.~\ref{sec:results} we present our novel results for the predictions of these models for
yielding in athermal quasistatic shear.  We start in
Sec.~\ref{sec:brittle} by showing that, beyond mean field, yielding is
always brittle in the limit of large system size. In contrast, for
small systems (or in mean field), yielding is brittle only for highly
annealed samples, and instead appears ductile for poor annealing. In
Sec.~\ref{sec:preFailure} we elucidate the physics of the
``pre-failure'' regime, in which small amounts of strain heterogeneity
slowly accumulate as a result of local plastic events that are
initially largely uncorrelated. Sec.~\ref{sec:failure} concerns the
catastrophic brittle failure event that follows. Here a  shear
band of highly correlated plastic events nucleates and spreads quickly
across the sample, leading to a discontinuous drop in the macroscopic
stress. In Sec.~\ref{sec:discussion}, we discuss the implications of our results for experiments and molecular simulations. Sec.~\ref{sec:conclusions} gives conclusions and perspectives
for future work.

Before describing the model to be simulated, we make the following
remarks to ensure that our terminology is unambiguous. 

First, we use
the term ``brittle'' to characterize abrupt yielding in which the
stress drops suddenly from its maximum value that pertains just before
yielding, to a final post-yield value, during a single event in which
the strain becomes localized along a system-spanning  shear
band. We call yielding ``ductile'' if the fall in stress happens more
gradually, or as a series of several smaller sudden drops, as a
function of the applied strain. This terminology is consistent with that as used in Ref.~\cite{ozawa2018random}. We note, however, that the term brittle is sometimes reserved to describe yielding via crack propagation in the absence of significant plasticity, whereas significant plasticity does occur inside a forming shear band. In places where we use the term ``brittle", therefore, readers concerned with this distinction may prefer to read it as ``quasi-brittle".

Second, we use the term ``shear
banding'' both in the sense typically adopted in the rheology
literature (to mean a sustained coexistence of macroscopic bands
flowing with different shear rates), and as used in the metallic glass
literature (to mean the formation of a highly localised band of deformation),
noting that these phenomena indeed also share many notable similarities. In
any place where the distinction is not clear from the context, we
shall clarify this explicitly. We note further that the formation of a shear band in studies with periodic boundary conditions (as the vast majority of theoretical works use) is likely in practice to be associated with the formation of a crack with new free surfaces, in an experimental system where the boundaries can move apart.

Third, we use the term ``nucleation''
to denote the onset of shear banding in metallic glasses for
consistency with the use of the term in that literature, rather than
necessarily to mean the thermally activated crossing of a barrier
leading to the formation of a new phase or pattern. (These phenomena
clearly nonetheless share notable similarities in their stochastic
nature.)

\section{Elastoplastic model}
\label{sec:model}

Because amorphous materials lack any long range crystalline order,
their behaviour cannot be understood in terms of (for example) the
motion of defects in a background lattice structure.  Instead, it has
been attributed to the generic presence of {\em structural disorder}
(think of a disordered arrangement of foam bubbles, for example), {\em
  metastability} (for foam bubbles to rearrange, the soap films must
first stretch, which typically incurs an energy cost much greater than
$k_{\rm B}T$), and {\em broken ergodicity} (because of these energy
barriers, the bubbles cannot rearrange in the absence of an externally
applied shear). 

A popular modelling approach is to distill these ingredients into a
mesoscopic description that bridges the gap in lengthscales between a material's
constituent microscopic substructures (foam bubbles, emulsion
droplets, etc), and its emergent macroscopic flow behaviour. The basic idea is to consider any macroscopic sample of material  to comprise many mesoscopic elastoplastic
elements, each representing a patch of several (e.g.) foam
bubbles. Each element is assumed to load
elastically in shear, in between intermittent local plastic events. In any plastic event, a
local energy barrier to particle rearrangements is surmounted, and
stress is  released.

In what follows we  shall simulate the behaviour in shear of a 2D lattice elastoplastic
model~\cite{ISI:000227459200003,nicolas2018deformation}, which has a
single elastoplastic element on each of $N\times N$ lattice
sites. Each element is assigned a local elastic shear strain $l$, a
corresponding local shear stress $kl$, and a local yield energy $E$.
In between local plastic yielding events, the strain of each element
affinely follows the globally applied strain $\gamma$, corresponding
to elastic loading with $\dot{l}=\gdot$. For any element, this elastic
loading continues until its strain energy reaches the yield threshold,
$\tfrac{1}{2}kl^2=E$. Beyond this threshold, the element yields
stochastically at a rate $\tau_0$.

When any element yields, it adopts a new local strain chosen at random
from a Gaussian distribution 
$l_{\rm w}$. The small value of $l_{\rm w}$ is not important to the
physics that follows and we set $l_{\rm w}=0.02$ throughout. (As
discussed below, we work in units in which the local yield strain
$l_{\rm c}\equiv
\sqrt{2E/k}=1$ for all elements.) This
yielding of an element at any lattice site models a local plastic
yielding event within the material: physically, this might correspond
to the sudden rearrangement of a few particles, say. Just after any
such local yielding event, the stress across the 2D lattice will not
be divergence free: {\it i.e.}, force balance will be violated. Force
balance is then immediately re-imposed via the Eshelby stress
propagator in 2D~\cite{picard2004elastic,eshelby1957determination}. In
this way, the change in plastic stress experienced locally at any
lattice site is propagated elastically to the surrounding medium. Specifically, the propagated elastic displacement field $\vecv{u}$ resulting from a plastic event that results in a local stress change $\tens{S}$ is given at wavevector $\vecv{q}$ in Fourier space by~\cite{picard2004elastic}
\be
\vecv{u}=\frac{1}{kq^2}(\tens{I}-\hat{\vecv{q}}\hat{\vecv{q}}).i\vecv{q}.\tens{S}.
\ee
The associated elastic stress follows as $\tfrac{k}{2}(\nabla\vecv{u}+\nabla\vecv{u}^{\rm T}$). As is common practice in the literature on elastoplastic models, we consider only the shear component of this.

We consider throughout the limit of quasistatic shear, $\gdot\tau_0 =
0$, such that local yielding in fact occurs instantaneously once
threshold is reached. Furthermore, in having assumed that the yielding
rate of any element is zero until the yield threshold is reached, we
are considering the athermal limit of zero temperature.  Accordingly,
the protocol studied here is that of ``athermal quasistatic shear''
(AQS), as considered widely in the literature on sheared amorphous
materials. We assume inertia to be negligible throughout and, as is standard in studies of elastoplastic models, consider the stress propagation that follows any local yielding event to be well approximated by that in a homogeneous elastic medium.

To initialise the sample prior to shear, we assign each element a
local shear strain $l$, chosen at random from a distribution
that has zero mean and standard deviation $l_0$. We then immediately impose
force balance across the lattice. This imposition of force balance
reduces the standard deviation of this initial distribution of local
strains by a factor equal to the square root of the sum over all
lattice sizes of the square of the Eshelby propagator. Throughout most of the paper, our numerical results will pertain to a Gaussian distribution of initial local strains. To demonstrate that our conclusions are robust with regards the shape of this distribution, however, we shall also perform calculations with a beta distribution of initial local strain values. To make a connection with some recent particle based simulations~\cite{patinet2016connecting,barbot2018local}, we shall furthermore consider a variant of our model in which all elements have zero initial local strain and instead have different local yield stress values, distributed with Weibull statistics. 

\begin{figure*}[!t]
\includegraphics[width=18.0cm]{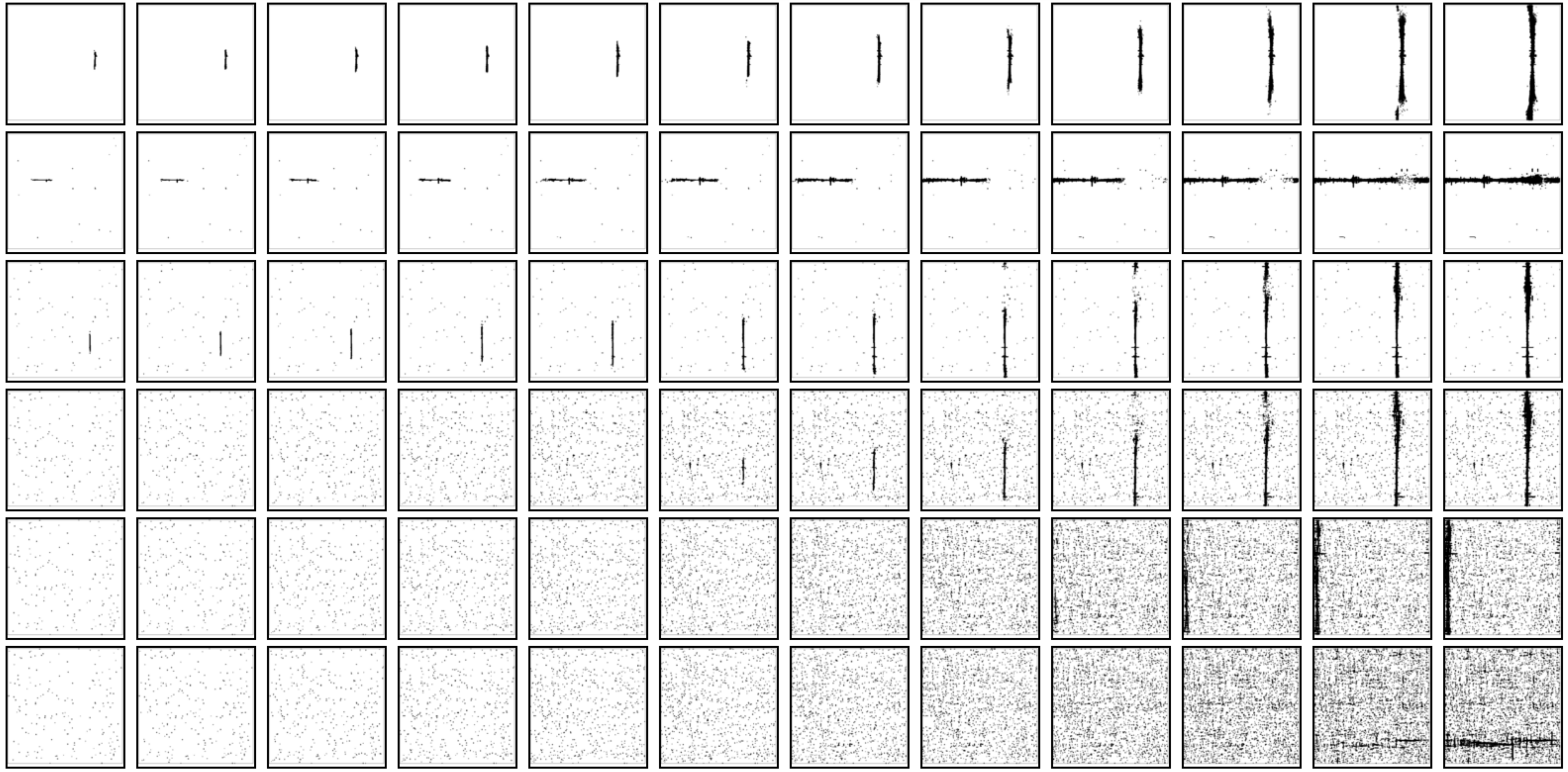}
\caption{Maps of sites that have yielded at least once (black) and never yielded (white) after $\rm{floor}(2.0^p)$ plastic events have occurred in total since the start of the deformation, with $p=7.5,9.0,\cdots 13.0$ from left to right for decreasing levels of initial sample annealing, $l_0=0.025,0.050,0.075,0.100,0.125,0.150$, from top to bottom. System size $N\times N=512\times 512$ in each map.}
\label{fig:crackMaps}
\end{figure*}

Small (resp. large) values of the parameter $l_0$ are taken to
correspond to highly (resp. poorly) annealed samples. In a physical
preparation protocol, a high degree of annealing could correspond
(say) to equilibration at a low temperature (only just above the glass
transition temperature), or to slow cooling from an initially high to
zero temperature, or to ageing for a long time at a low temperature
(below the glass transition temperature), before quenching to zero
temperature. (Conversely, poor annealing would correspond to
equilibration at a higher temperature before quench, faster cooling,
or a shorter ageing time.) We do not model any of these annealing
processes explicitly, but rather take the value of the single
parameter $l_0$ as a proxy for the level of annealing, as is standard
practice in studies of elastoplastic models. 

So far, we have described the `full', spatially aware elastoplastic
model, with explicit force balance via an Eshelby propagator, in two
spatial dimensions (2D). To simulate a mean field version of this
model, we randomly shuffle the location of all elements on the lattice
after each yielding event and the subsequent force rebalancing
step. This is not the same as simply not implementing force balance,
because the propagation of stress during force balance changes the
distribution of local strains.

At any strain step, our numerical algorithm is as follows. First, we
query how much elastic strain $\Delta\gamma$ must be applied in order
to take the least stable element just above its yielding
threshold. This amount of strain is then applied to every element on
the lattice, $l_i\to l_i+\Delta\gamma$, and the global strain variable
incremented by the same amount, $\gamma\to\gamma+\Delta\gamma$. The
least unstable element, now just above threshold, is then
yielded. Force balance is then reimposed across the lattice. This may
lead to other elements then exceeding their local thresholds. Any such
elements are then yielded, and force balance is reimposed again. This
process is repeated iteratively until no elements are left above yield after
rebalancing. We then proceed to the next strain step. If, in any
strain step, the amount of elastic strain needed to take the least
stable element above threshold falls below $\Delta\gamma=0.001$, the
strain increment is set to this minimal value.

Throughout we use units in which the elastic constant $k=1$, and adopt
the same yield energy threshold $E$ for for all elements, chosen such
that the yield strain of each element $l_{\rm c}\equiv
\sqrt{2E/k}=1$, as noted above.  
The
physical parameters that remain to be explored are then $l_0$, which
sets the degree of initial sample annealing prior to shear (recall
that small values of $l_0$ correspond to highly annealed samples), and
the size of the sample of material being sheared, as set by the linear
size $N$ of the two dimensional $N\times N$ lattice.

The macroscopic stress is defined as the average across the lattice of
the local elemental stresses:
\be
\Sigma=\frac{k}{N^2}\sum_{i=1}^{N^2}l_i.
\ee

\section{Results}
\label{sec:results}

The basic physics that we shall consider is summarised in
Fig.~\ref{fig:crackMaps}, which shows a collection of state snapshots,
each pertaining to a system of size $N=512$ for the model with full 2D
force balance. Each snapshot shows as a black dot the lattice sites
that have undergone at least one plastic yielding event since shearing
commenced. Unyielded sites are shown in white. Rows downwards
correspond to progressively less well annealed samples.  Columns left
to right show the system's state after an increasing number of plastic
events have occurred since shearing started.

At early strains after shearing starts, only a few plastic events
occur, scattered throughout the sample in a largely uncorrelated
way. These lead to a low level of strain heterogeneity slowly
accumulating across the sample. We shall call this the ``pre-failure''
regime. This regime then terminates (for large samples or strong
initial annealing) in a catastrophic brittle failure event, in which
plastic events spread quickly across the sample in a highly correlated
way, in the form of a shear band, and the macroscopic stress
signal drops discontinuously.

\subsection{Yielding is always brittle for large system sizes}
\label{sec:brittle}

\begin{figure}[!t]
\includegraphics[width=1.0\columnwidth]{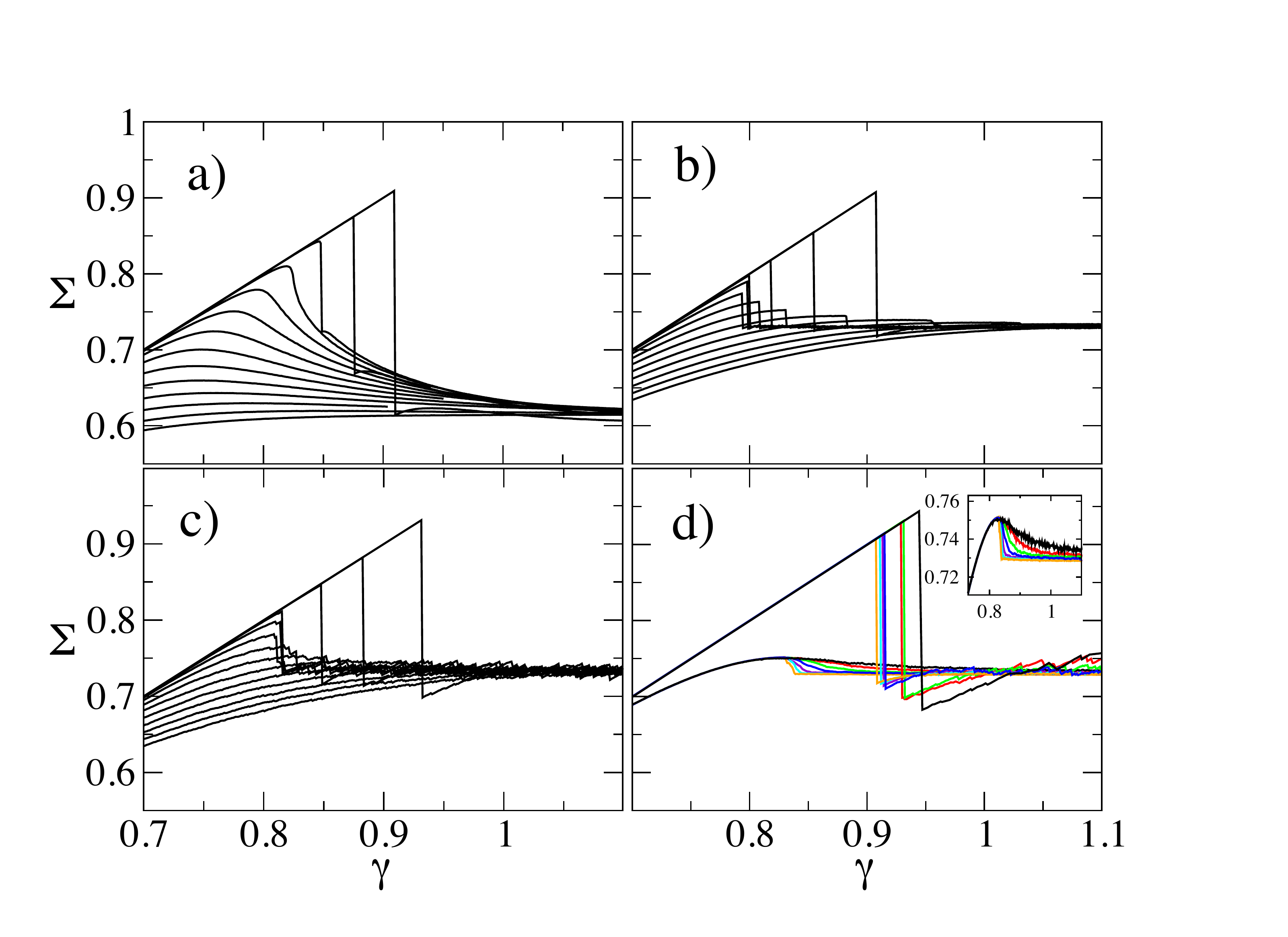}
\caption{Strain-strain curves of the lattice elastoplastic model. {\bf (a)} Mean field model for a lattice size $N\times N = 4096\times 4096$, with decreasing levels of sample annealing in curves downwards: 
$l_0=0.025, 0.050, 0.075\cdots 0.350$.  {\bf (b)} Full model with
explicit force balance, for the same parameters as in (a). {\bf (c)}
Full model, but now for a smaller lattice size $256 \times 256$. {\bf
(d)} Full model for two different annealing levels $l_0=0.025$ (upper
curves) and $l_0=0.200$ (lower curves) for system sizes $N=64, 128,
256, 512, 1024, 2048, 4096$ in black, red, green, blue, violet, cyan and orange  respectively.  Inset shows a zoom of the curves for $l_0=0.200$. The curves for $l_0=0.200$ are  at any $N$ averaged over $8\times 4096/N$ runs, each with a different random number seed.}
\label{fig:stressStrain}
\end{figure}
\begin{figure}[!t]
\includegraphics[width=1.0\columnwidth]{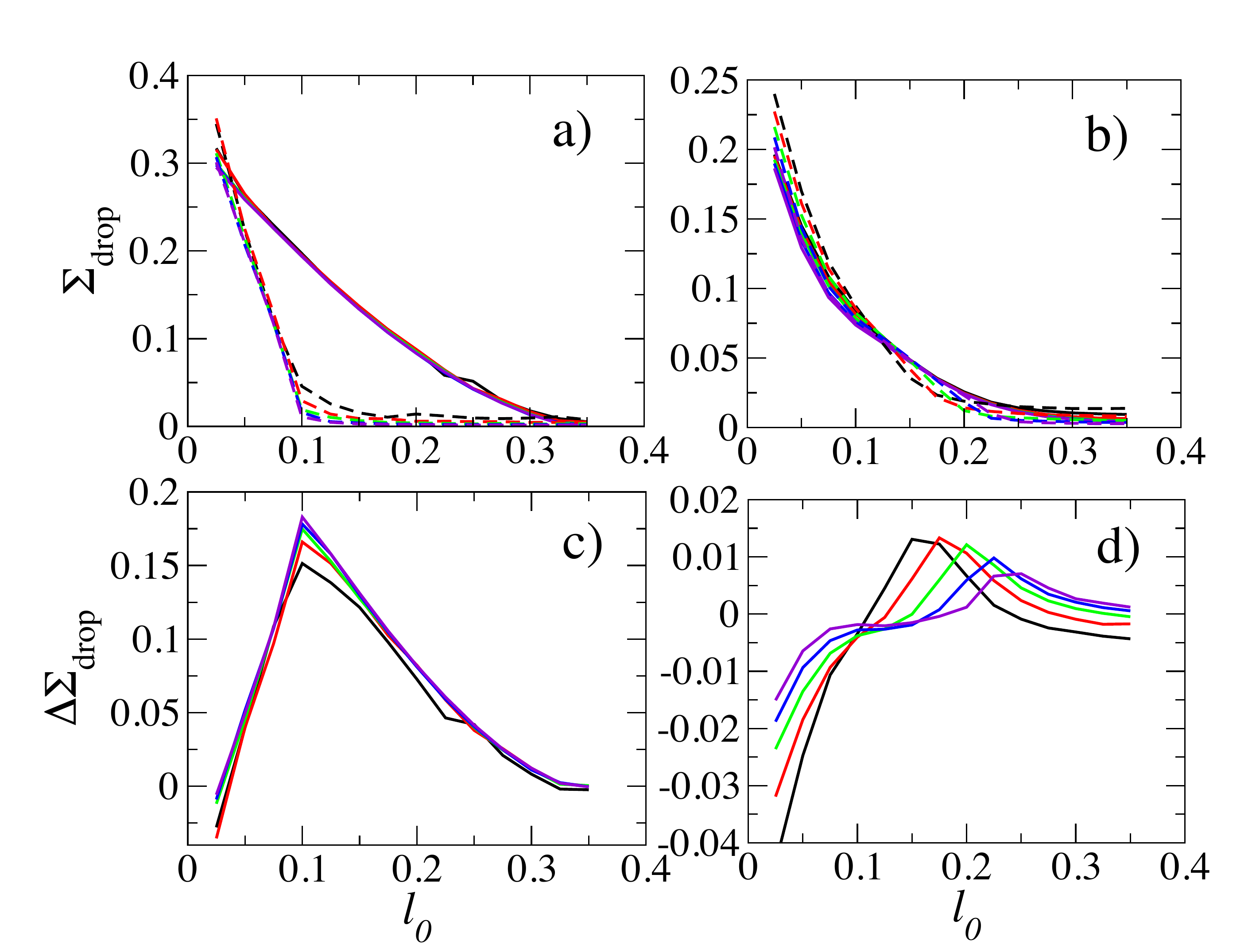}
\caption{Drop in stress-strain curves  as a function of the annealing parameter $l_0$, with smaller values of $l_0$ corresponding to better annealed samples. {\bf a+c)} In the mean field model. {\bf b+d)} In the full model with force balance. {\bf a+b)} Solid lines: total stress drop (maximum stress minus steady state stress). Dashed lines: discontinuous part of the stress drop (maximum stress drop in any single strain increment). {\bf c+d)} Difference between total stress drop and discontinuous part of stress drop. Sample sizes $N=128, 256, 512, 1024, 2048$ in black, red, green, blue, violet respectively. For each value of $N$, drops are averaged over $8\times 4096/N$ runs, each run with a different value of the random number seed.}
\label{fig:drops}
\end{figure}

We shall start by confirming that the scenario of Refs.~\cite{ISI:000436245000061,popovic2018elastoplastic} also holds in the mean field version
of the elastoplastic model simulated here, before then showing that it
does not apply beyond mean field. See the stress-strain curves in
Fig.~\ref{fig:stressStrain}a). These show a discontinuous stress drop
for highly annealed samples (small values of $l_0$), but a continuous
stress drop for poor annealing (larger $l_0$). (For the largest values
of $l_0$, the stress rises monotonically with strain.)  As sketched in
Fig.~2b) of Ref.~\cite{popovic2018elastoplastic}, the discontinuous
stress drop for strong annealing can be understood (in mean field) to
stem from an underlying stress-strain curve that has an overhang, with
the stress falling discontinuously from the curve's upper to lower
branch once the overhang is reached.

The distinction between the regime of discontinuous stress drop for
highly annealed samples and of continuous stress drop for poorly
annealed samples is further explored in Fig.~\ref{fig:drops}.  We
define the net stress drop during a shear simulation, for any given
level of initial sample annealing, to be the difference between the
maximum stress (maximised over all strains) and the (statistically)
steady state stress approached as strain $\gamma\to\infty$. We report
this net stress drop as a function of the annealing parameter $l_0$
via the solid lines in Fig.~\ref{fig:drops}a). We now further
consider only the discontinuous part of this stress drop, defined as
the maximum stress drop in any single strain increment. We report this
quantity via the dashed lines in Fig.~\ref{fig:drops}a). The
difference between these two drops is plotted in
Fig.~\ref{fig:drops}c). As can be seen, a discontinuous stress drop
exists for high levels of sample annealing, with $l_0<l_0^*\approx
0.1$. Its amplitude then falls continuously to zero as $l_0$ rises
towards $l_0^*$.  For poorly annealed samples ($l_0>l_0^*$), there is
no discontinuous stress drop, at least for large systems. This
behaviour furthermore shows only a rather weak dependence on system
size, $N$.

So far, then, we have confirmed the scenario reported in earlier mean
field studies of elastoplastic
models~\cite{,popovic2018elastoplastic}, with brittle
yielding for strong annealing ($l_0<l_0^*$), ductile yielding for poor
annealing ($l_0>l_0^*$), and a ``random critical point'' separating
these at $l_0=l_0^*$.

We now move beyond mean field and propose that the route to reconciling the apparent
contradiction between
Refs.~\cite{ISI:000436245000061,popovic2018elastoplastic} (brittle
yielding for high annealing and ductile yielding for poor annealing),
and Ref.~\cite{barlow2020ductile} (brittle yielding for all levels of
annealing, provided a stress drop indeed occurs), as discussed in Sec.~\ref{sec:intro} above, lies in a careful
study of the size of the sample being sheared. 

To demonstrate this, we
shall simulate our 2D elastoplastic model of $N\times N$ elastoplastic
elements with Eshelby force balance, for a previously unprecedented
range of system sizes from $N=64$ to $4096$. For small $N$ we shall
recover the observation of
Refs.~\cite{ISI:000436245000061,popovic2018elastoplastic}, of brittle
yielding for high annealing and ductile yielding for poor
annealing. This distinction is however not governed by a critical
point in our simulations. For large $N$ we shall recover the scenario
of Ref.~\cite{barlow2020ductile}, with brittle yielding for all levels
of annealing, provided a stress drop indeed exists. In this way, we
suggest that the scenario found in the earlier studies of 2D
elastoplastic
models~\cite{ISI:000436245000061,popovic2018elastoplastic}, which
apparently confirmed the random critical point found in mean field,
may be a finite size effect.

The reconciliation just described is substantiated in
Figs.~\ref{fig:stressStrain} and~\ref{fig:drops}, as follows.
Fig.~\ref{fig:stressStrain}b) shows stress-strain curves computed
within the 2D elastoplastic model with full force balance, for a large
system size $N\times N=4096\times 4096$. Curves with decreasing
maximum stress values correspond to decreasing levels of annealing. As
can be seen, the stress drop is discontinuous not only for the highly
annealed samples, but also (or very nearly so, for this $N=4096$) even
for poorly annealed samples. In contrast, simulating the same
annealing levels for a small system size $N=256$ in
Fig.~\ref{fig:stressStrain}c) shows a convincingly discontinuous
single stress drop only for strong annealing. In poorly annealed
samples, the stress instead declines continuously, when averaged over
many runs with different random number seeds. (In any single run, it decreases
via a cumulative series of smaller discontinuous
drops.) Fig.~\ref{fig:stressStrain}d) shows two collections of curves:
the upper collection for strong annealing, and the lower collection
for weak annealing, with each collection showing a full sequence of
system sizes, $N=64,128,256\cdots 4096$. For the highly annealed case,
the stress drop is discontinuous even for small $N$. For the poorly
annealed case, the stress drop is gradual and continuous for small
$N$, but becomes increasingly steeper with increasing $N$.

This is explored further in Fig.~\ref{fig:drops}, in which panel b)
shows as solid lines the net stress drop (from the maximum stress
value to the steady state) as a function of the annealing parameter
$l_0$, averaged over many runs at each $l_0$. The discontinuous part
of the stress drop, also averaged over many runs at each $l_0$, is
shown as dashed lines. 
As can be seen, the discontinuous stress drop persists across the full
range of $l_0$ in this non-mean field system, even for poorly annealed
samples (high $l_0$). This is to be contrasted with the vanishing of
the discontinuous stress drop at the critical point $l_0=l_0^*\approx
0.1$ in the mean field model.  We therefore conclude that the critical
point of the mean field model does not inform the yielding behaviour
of the non-mean field model.

Subtracting the discontinuous part of the stress drop from the net
stress drop for the non-mean field model in panel d), we see that the
difference between the net and the discontinuous part of the stress
drops shifts progressively to zero as the limit of infinite system
size $N\to\infty$ is approached. In the limit $N\to\infty$, therefore,
we expect the entire stress drop to occur discontinuously in a single
strain increment, however poorly annealed the sample, corresponding to
brittle yielding.

For very poorly annealed samples, with a broad initial distribution of
local strains, the stress rises monotonically with strain, with no
stress overshoot or drop. Indeed, in this case, significant plasticity
occurs early in any simulation, because elements in the positive-$l$
wing of the initial strain distribution exceed their local threshold
already at small strains. This is true even for large $N$ and clearly
would be described as ductile rather than brittle behaviour. However,
whether it would be described as {\em yielding} is a moot point,
because yielding describes the progression with increasing strain from
an initially elastic state to a finally plastic one.

\subsubsection*{Robustness to shape of initial local strain distribution}

So far, the elements in our elastoplastic model have all had the same local yield strain, $l_{\rm c}$, but different initial local strain values drawn from a Gaussian distribution.  From this it follows that the values of the initial stress needed to induce a local yielding event -- the   ``stress-to-yield'' values -- are also drawn from a Gaussian distribution. Recent atomistic simulations~\cite{patinet2016connecting,barbot2018local} have however indicated that the distribution of the stress-to-yield values, approaches zero as a power law, in the limit of small stress-to-yield. In order to test the robustness of our predictions to the shape of the initial distribution of stress-to-yield values, we now consider two alternative initial conditions, aimed at capturing the statistics reported in Refs.~\cite{patinet2016connecting,barbot2018local}.

The first of these keeps the yield stress values fixed across sites, at $l_{\rm c} = 1$, but replaces the Gaussian distribution of initial strain values with a $\beta$-distribution. This has a probability density function given by
\begin{equation} 
\label{eq:pdf_beta}
  f(l) = \frac{\Gamma(2\alpha)}{2\Gamma^2(\alpha)} \left(\frac{1+l}{2} \right)^{\alpha-1}\left(\frac{1-l}{2}\right)^{\alpha -1},
\end{equation}
for $-1 \leq l \leq 1$, with $f(l) = 0$ outside this region. Here $\Gamma$ is the standard Gamma function. The $\beta-$distribution  has zero mean, is symmetric about this mean, and decays (for $\alpha>1$) towards the yield strain $l_{\rm c}=1$ as a power law with exponent $\alpha-1$. This results in a power law decay of the distribution of stress-to-yield values, close to yield, as seen in the particle simulations~\cite{patinet2016connecting,barbot2018local}.  After drawing the initial local strain values from this distribution we then impose force balance across the material before shear is applied. The other aspects of our simulation remain as described in Section 2. 
\begin{figure}[!t]
\includegraphics[width=1.0\columnwidth]{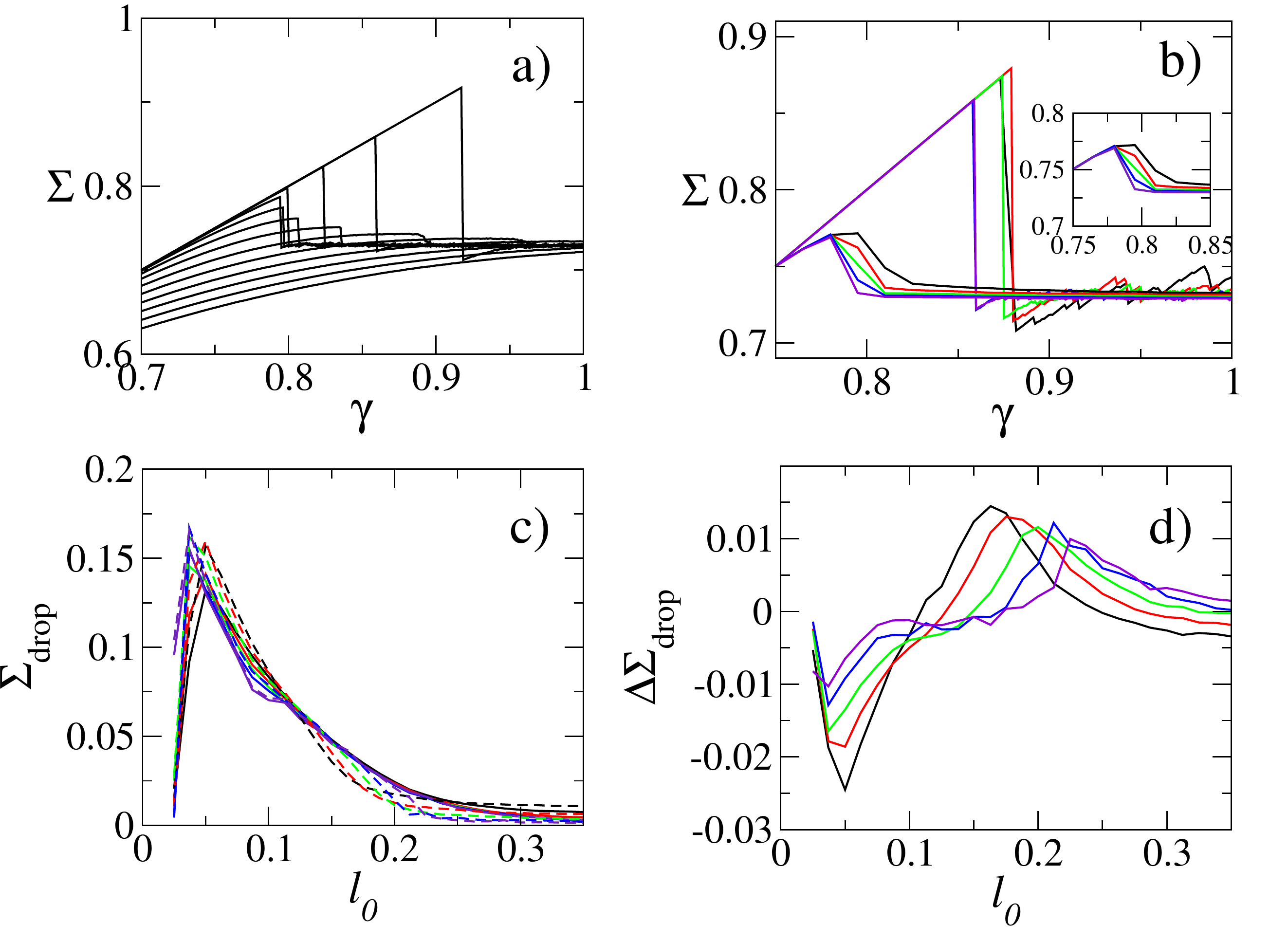}
\caption{Strain-strain and stress drop curves of the lattice elastoplastic model with the initial stresses distributed according to a beta distribution as in Eq.~(\ref{eq:pdf_beta}). (a) Stress-strain curves for a lattice size $N \times N = 2048 \times 2048$, with decreasing levels of sample annealing in curves downwards: $l_0 = 0.025, 0.050, 0.075 \cdots 0.350$. (b) Stress-strain curves for two different annealing levels $l_0 = 0.025$ (upper curves) and $l_0 = 0.1$ (lower curves) for system sizes $N = 128, 256, 512, 1024, 2048$ in black, red, green, blue, and violet respectively. Inset shows a zoom of the curves for $l_0 = 0.1$. The curves for $l_0 = 0.1$ are at any N averaged over $8 \times 4096/N$ runs, each with a different random number seed. (c) Solid lines: total stress drop (maximum stress minus steady state stress). Dashed lines: discontinuous part of the stress drop (maximum stress drop in any single strain increment). (d) Difference between total stress drop and discontinuous part of stress drop. Sample sizes $N = 128, 256, 512, 1024, 2048$ in black, red, green, blue, violet respectively. For each value of $N$, drops are averaged over $8 \times 4096/N$ runs, each run with a different value of the random number seed.}
\label{fig:beta_figure}
\end{figure}

As usual, we model the degree of annealing by the variance $l_0^2$ of the distribution of initial local strain values: a well-annealed sample has a small $l_0$, and a poorly annealed sample a large $l_0$. The exponent $\alpha$  of the $\beta-$ distribution is related to its variance via:
\begin{equation}
  \alpha = \frac{1}{2}\left(\frac{1}{l_0^2} -1\right).
\end{equation}
The condition $\alpha > 1$, needed for the distribution to decay near threshold, holds as long as $l_0 < 1/\sqrt{3} \approx 0.5774$. 

\begin{figure}[!t]
\includegraphics[width=1.0\columnwidth]{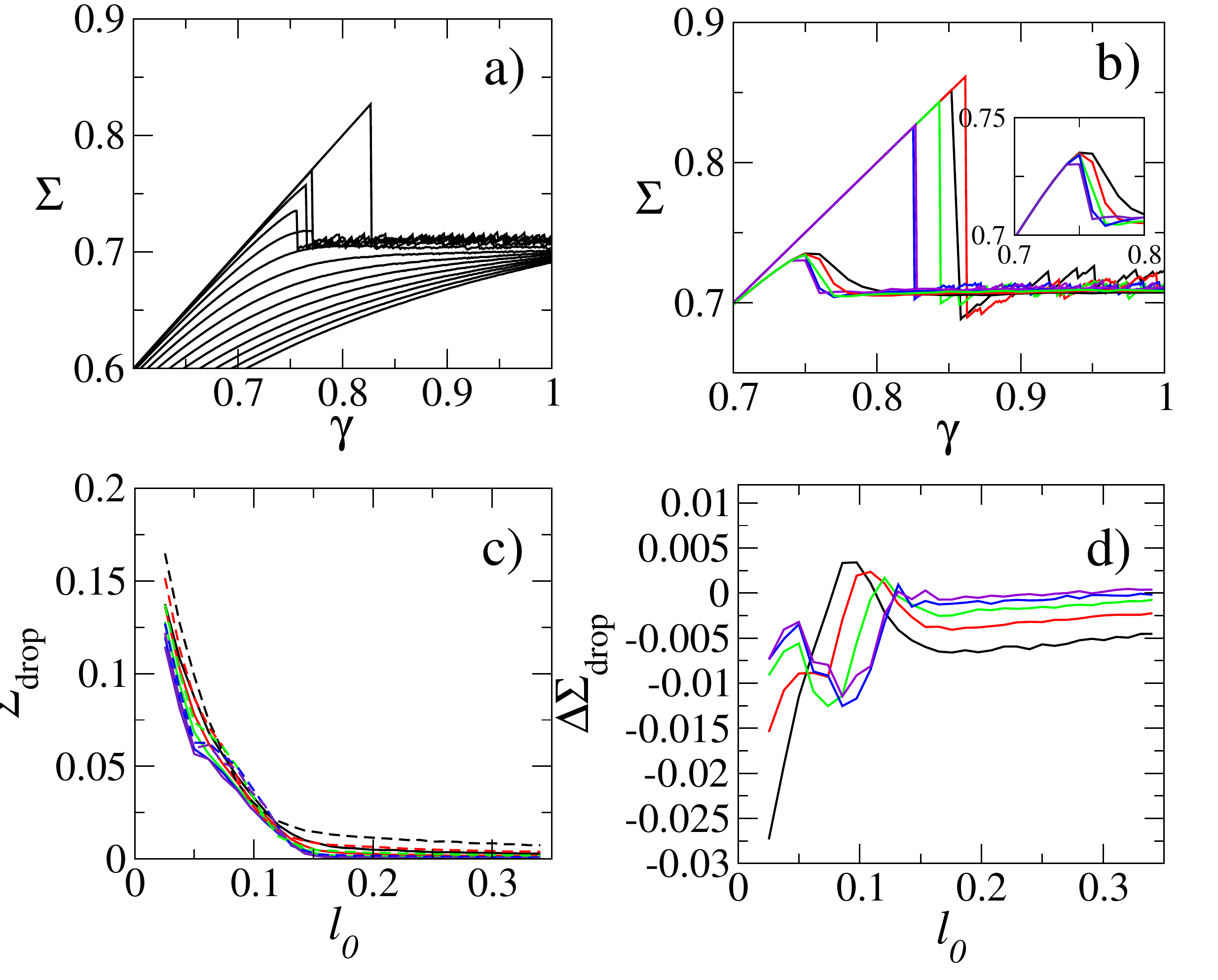}
\caption{Strain-strain and stress drop curves of the lattice elastoplastic model with fixed initial stress and the yield stresses distributed according to a Weibull distribution as in Eq.~(\ref{eq:pdf_weibull}).  Panels a)-d) have the same parameter values as their direct counterparts in Fig.~\ref{fig:beta_figure}.}
\label{fig:weibull_figure}
\end{figure}

Stress-strain curves for different levels of annealing for this beta distribution, with $l_0$ values chosen to match the variances considered for a Gaussian distribution in Fig.~\ref{fig:stressStrain}(a,b), are shown in Fig.~\ref{fig:beta_figure}(a,b). Consistent with our observations in Fig.~\ref{fig:stressStrain}, we find a discontinuous stress drop for highly annealed samples for all sample sizes. For poorly annealed samples, we instead find a continuous stress drop for small and moderate system sizes, with the drop then tending to a discontinuous one in the limit of large system size. This is further explored in Fig.~\ref{fig:beta_figure}(c), which shows the total and (separately) the discontinuous part of the stress drop. The difference between these are shown in  Fig.~\ref{fig:beta_figure}(d), and found to decay slowly as system size increases. In the limit of infinite system size, therefore, the entire stress drop is predicted to be discontinuous, corresponding to brittle yielding. 

In our second alternative to a Gaussian distribution of initial local strains, we remove the randomness from the initial local strains, taking the initial strain of each element to be zero. We instead now adopt a distribution of initial local yield strain values, $l_{\rm c}$. For these we assume a Weibull distribution, which represents a generalisation of the exponential distribution, and which fits the form of the stress-to-yield curves obtained in Refs.~\cite{patinet2016connecting,barbot2018local}, near threshold. The probability density function is 
\begin{equation} \label{eq:pdf_weibull}
  f(l) = \frac{k}{\lambda} \left(\frac{l}{\lambda}\right)^{k-1} e^{-(l/\lambda)^k}.
\end{equation}
This has two parameters: a scale factor $\lambda$ and an exponent $k \geq 1$, with the exponential distribution corresponding to $k=1$. The scale parameter simply determines the units of stress. We choose to work in units such that the mean of the distribution is equal to 1, which is equivalent to choosing
\begin{equation}
  \lambda = \frac{1}{\Gamma(1+1/k)}.
\end{equation}
This leaves us with the single parameter, $k$, which sets the exponent of the power law decay $l^{k-1}$ of the stress-to-yield values, close to zero.  It is related to the variance $l_0^2$ of the distribution by
\begin{equation}
\label{eqn:invert}
  l_0^2 = \frac{\Gamma(1+2/k)}{\Gamma^2(1+1/k)} - 1.
\end{equation}
Accordingly large values of $k$ correspond to small $l_0$ and so to well annealed samples. We consider values of $l_0$ so as to match the values of the variance used previously for the Gaussian and $\beta$ distributions. At any $l_0$, we invert Eqn.~\ref{eqn:invert} numerically to find the corresponding value of $k$.
To ensure that the final state of steady flow is independent of the initial condition, after any element's first local yielding event we choose its new yield strain from a Weibull distribution of unit mean and fixed variance $l_w^2 = 0.25$. Otherwise our algorithm is as described in Section 2.

The resulting stress-strain curves are shown in Fig.~\ref{fig:weibull_figure}(a,b). The net stress drop and discontinuous part of the stress drop are shown in Fig.~\ref{fig:weibull_figure}(c,d). All the key features found previously for a Gaussian distribution are preserved, with brittle failure for highly annealed samples, and a slow progression from ductile to brittle failure with increasing system size for poorly annealed samples.   

Having shown the same qualitative behaviour for three different shapes of the distribution of initial stress-to-yield values, we suggest that our physical conclusions are robust with respect to that shape.

\subsection{``Pre-failure'' stage: slow accumulation of strain heterogeneity}
\label{sec:preFailure}

Returning to Fig.~\ref{fig:crackMaps}, we recall that yielding
comprises two regimes. In the first regime, plastic events arise
sporadically in a way that is (at least initially) relatively
spatially uncorrelated. For highly annealed samples (upper rows in
Fig.~\ref{fig:crackMaps}), these plastic events occur very
infrequently; for poor annealing (lower rows), they are more
frequent. In the second regime, a macroscopic shear band spreads quickly
across the sample and the sample fails suddenly. In this section we aim to understand the first
regime, before sudden failure occurs.

In this first regime, the occurrence of plastic events throughout the
sample leads to a progressive accumulation of strain
heterogeneity. This can be understood most easily within a simplified
one-dimensional description in which we consider our $N\times N$
lattice in $x$ and $y$ to comprise an array of $N$ streamlines
arranged across the $y$ direction, along each of which we project down
the $x$ coordinate by integrating over the $N$ elements along $x$. (We
could equally instead consider an array of $N$ streamlines across $x$,
integrating over the $N$ elements along $y$.)

It is worth noting that this equivalence between $x$ and $y$ holds on account of the $\pi/2$ rotational symmetry of the Eshelby propagator, in tandem with the fact that,
in common with most numerical studies of elastoplastic
models, we have ignored a term that would advect elements with the
shear flow. Had we included advection, with a flow direction $x$ and
flow-gradient direction $y$, we anticipate that the strain
heterogeneity discussed in this pre-failure regime would arise across
the flow-gradient direction $y$, with the system's state remaining
relatively invariant in $x$: any heterogeneity with wavevector
along the $x$ direction would be advected to finally have its wavevector
indeed along $y$.

Consider then (in our model in which $x$ and $y$ are equivalent) a
plastic event having just occurred on a given streamline at some
$y$. This causes a loss of stress on that streamline of $1/N$. (The
streamline stress is defined as the average of the local elemental
stresses along that streamline.) This streamline is therefore now out
of force balance with the others: in a balanced system, the stress is
the same across all streamlines. In the force balance step that
immediately follows, therefore, the streamline on which the plastic
event just occurred must strain forwards slightly to recover the same
stress as all the other streamlines. Every streamline furthermore then
strains backwards (by a lesser amount) to respect the fact that some
stress has been lost from the system in the plastic event. The physics
just described in words is what indeed results after imposing a full
Eshelby 2D propagator, and then projecting down the result to one
dimension.

Accordingly, after a given strain $\gamma$ has been applied (on
spatial average) across sample as a whole, any streamline of constant
$y$ (or of constant $x$) in Fig.~\ref{fig:crackMaps} that has
experienced a higher number of plastic events than the average number
across the whole sample will have strained forwards a little more than
the average $\gamma$. Those that have experienced statistically fewer
events will have strained forwards a little less than $\gamma$.

\begin{figure}[!t]
\includegraphics[width=1.0\columnwidth]{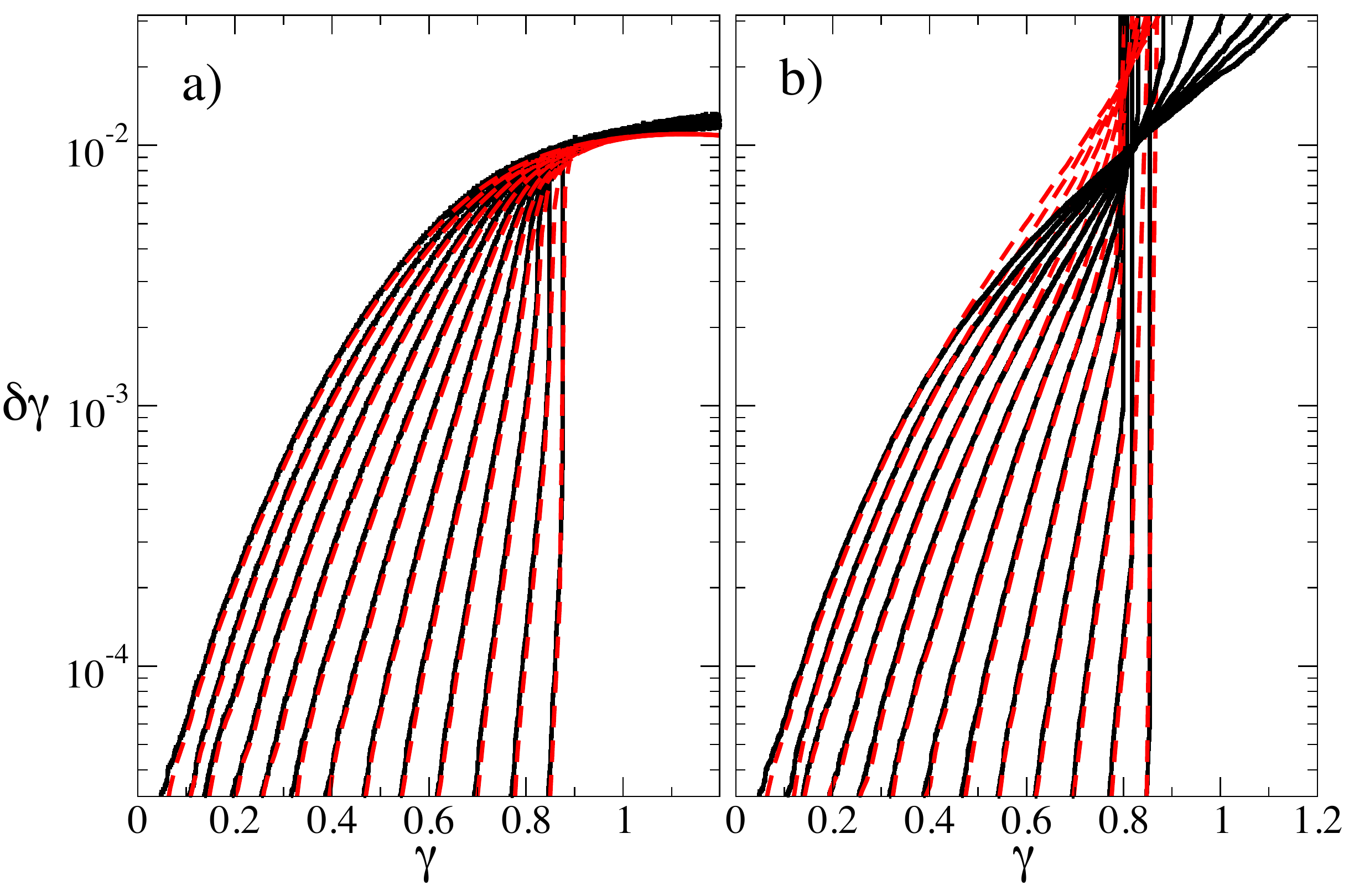}
\caption{{\bf a)} Black solid lines: Growth of the small standard deviation of strain across the sample  $\delta \gamma$ as a function of overall imposed strain $\gamma$ in the prefailure regime, measured from simulations of a mean field elastoplastic model with a lattice size $N\times N = 4096\times 4096$ for decreasing levels of sample annealing in curves leftwards: 
$l_0=0.050, 0.075, 0.100\cdots 0.350$.  Red dashed lines: analytical prediction for this quantity according Eqn.~\ref{eqn:preFailureMF}. {\bf b)} As in panel a), but now  with full force balance, and with the analytical prediction (red dashed lines) given by Eqn.~\ref{eqn:preFailure}. (Once any solid line turns sharply upwards, catastrophic failure has set in.) }
\label{fig:preFailure}
\end{figure}

Let us therefore now define the strain heterogeneity,
$\delta\gamma(\gamma)$, as the standard deviation of the strain across
the $y$ direction, having first integrated across $x$.
This is plotted as a function of the average
strain $\gamma$ applied to the sample as a whole since shearing
commenced in Fig.~\ref{fig:preFailure}.  Results are shown for the
mean-field model in panel a) and for the 2D lattice model with force
balance in b). The solid lines are the strain heterogeneity measured
from our simulations. The dashed lines are the results of the
analytical calculation that we shall now present in order to
understand these simulation results.

Let us start by considering the distribution of local strains as first
initialised, denoted $\tilde{P}(l,t=0)$. Recall that in most of our numerical results this is a
Gaussian of width $l_0$. (Robustness against switching to a beta or Weibull distribution having been demonstrated above. Indeed, the analytical arguments that follow below do not depend upon a particular choice for the shape of the distribution.) As described in Sec.~\ref{sec:model}, we then
immediately impose force balance. This narrows the
width of the Gaussian by a factor equal to the square root of the sum
across all lattice sites of the square of the 2D Eshelby
propagator. We denote this re-normalised distribution by
$P(l,t=0)$. It represents the state of the system immediately before
shearing commences.

In what follows, consider the evolution of the probability distribution
$P(l,t)$ as a function of the time $t$ or (almost) equivalently the
accumulating strain $\gamma=\gdot t$ applied to the lattice as a
whole. We also consider the associated macroscopic stress
$\Sigma=\int dl\, l\, P(l,\gamma)$ and the strain variable itself,
$\gamma$. Our strategy will be to examine how these quantities evolve
in a perfectly homogeneous ``base state'' in which the strain remains
everywhere equal to $\gamma$, with no strain heterogeneity, and then
to elucidate via a linear stability analysis the fate of small
amplitude heterogeneous perturbations to this base state.

An evolution equation for the probability density function of local strains, $P(l,t)$,
can be written as follows (for the moment suppressing any space
dependencies in our notation):
\be
\label{eqn:evolution}
\partial_tP(l,t)+\gdot \partial_lP = -r(l)P + Y(t)P_{\rm w}(l).
\ee
Here $r(l)$ is the yielding rate of an element with local strain
$l$. Recall that $r=0$ for $|l|<1$ and $r=1$ for $|l|>1$.  The average
rate of plastic yielding across the lattice $Y(t)=\int dl\, r(l)\,
P$. The second term on the LHS of Eqn.~\ref{eqn:evolution} accounts
for the elastic loading of elements with strain rate $\gdot$. The
first term on the RHS accounts for their ``death'' due to plastic
yielding, and the second for their ``rebirth'' into new traps, with a
post-yielding local strain selected from a distribution $P_{\rm w}(l)$ of
zero mean and small width $l_{\rm w}$. 

Pre-multiplying Eqn.~\ref{eqn:evolution} across by $l$ and averaging
over $l$ then gives an evolution equation for the macroscopic stress $\Sigma$:
\begin{eqnarray}
\label{eqn:stressEvolution}
\dot{\Sigma}&=&\gdot-\int dl\,l\,r(l)P,\nonumber\\ 
            &=&\gdot\left[1-P(l,t)|_{l=1}\right].
\end{eqnarray}
In moving from the first line to the second, we have recognised that
in the limit of quasistatic shear explored in this work, $\gdot\to 0$,
each element will yield instantaneously once it reaches its threshold
$l=1$.

We now switch independent variables from the time $t$ since shearing
commenced at $t=0$ to the strain $\gamma=\gdot t$ applied to the
sample as a whole since $t=0$, with $\gdot$ the corresponding strain
rate. We then have
\begin{equation}
\label{eqn:onStrain}
\frac{d\Sigma}{d\gamma}=1-P(l,\gamma)|_{l=1}.
\end{equation}

In the early stages of the deformation, before significant numbers of elements have yielded, the distribution $P(l,\gamma)$ can be written to good approximation as
\begin{eqnarray}
\label{eqn:shifting}
P(l,\gamma)&=&P(l-\gamma,\gamma=0)\;\;\textrm{for}\; l<1,\nonumber\\
P(l,\gamma)&=&0\;\;\;\;\;\;\;\;\;\;\;\;\;\;\;\;\;\;\;\;\;\;\;\textrm{for}\; l>1.
\end{eqnarray}
This encodes the fact that the stress distribution shifts affinely
upwards with elastic loading, up to the yield threshold $l=1$ at which
elements ``die'', and remains relatively unperturbed by the
``rebirth'' of elements via plastic rearrangements until significant
numbers of elements have yielded, in the later stages of deformation.

Our equation of motion, now specialised to the early stages of
quasistatic shear, can therefore be written:
\begin{equation}
\label{eqn:stressEvolution2}
\frac{d\Sigma}{d\gamma}=1-P(l-\gamma,\gamma=0)_{l=1}.
\end{equation}

As written,  Eqns.~\ref{eqn:evolution}
to~\ref{eqn:stressEvolution2} appear to neglect the effect of force
rebalancing. This can be included by considering a shear strain
$\gamma$ that is now not just the homogeneous strain averaged over the
sample as a whole, but an effective heterogeneous strain that also
accounts for any shears that arise internally during force balancing,
and which lead also to a heterogeneous distribution $P$.

Let us now incorporate this in our simplified 1D approach. As noted
above, our tactic is to consider an initially homogeneous ``base
state'' in which the strain remains everywhere equal to the globally
applied strain $\gamma$, with no heterogeneity; and then to study the
fate of small heterogeneous perturbations to this base
state. Accordingly, we write:
\begin{eqnarray}
\Sigma(y,t)&=&\Sigma_0(t)+\sum_k\delta\Sigma_k(t)\exp(iky),\nonumber\\
\gamma(y,t)&=&\gamma_0+\sum_k\delta\gamma_k(t)\exp(iky),\nonumber\\
P(l,y,t)&=&P_0(l,t)+\sum_k\delta P_k(l,t)\exp(iky),
\end{eqnarray}
in which $(\Sigma_0,\gamma_0,P_0)$ denote the base state and
$(\delta\Sigma_k,\delta\gamma_k,\delta P_k)$ the small
perturbations. Although we have decomposed the perturbations into
Fourier modes, each of wavevector $k$, the governing equation of
motion in fact contains no space-differential operators. (This is a
property of the Eshelby propagator projected down to
1D~\cite{picard2004elastic}.) Accordingly, each $k-$mode will behave
in the same way and we now drop the subscript $k$.

Substituting the base state plus small perturbations into
Eqn.~\ref{eqn:stressEvolution2}, expanding in powers of the small
perturbations and retaining only terms of first order then gives a
linearised equation of motion for the perturbations:
\begin{eqnarray}
\label{eqn:linearised}
\frac{d\delta\Sigma}{d\gamma}&=&\frac{d\delta\gamma}{d\gamma}\left[1-P_0(l-\gamma,\gamma=0)_{l=1}\right]\nonumber\\
             & &-\delta\gamma\, \partial_\gamma P(l-\gamma,\gamma=0)|_{l=1}\nonumber\\
             & &-\delta P(l-\gamma,\gamma=0)_{l=1}.
\end{eqnarray}

Force balance dictates that the stress must remain uniform across
streamlines, and accordingly that $\delta\Sigma=0$. Inserting this
into Eqn.~\ref{eqn:linearised}, and rearranging, gives:
\be
\label{eqn:linearised1}
\frac{d\delta\gamma}{d\gamma}=f(\gamma)\delta\gamma+\delta g(\gamma),
\ee
in which
\be
\label{eqn:f}
f=\partial_\gamma P(l-\gamma,\gamma=0)|_{l=1}/\left[1-P_0(l-\gamma,\gamma=0)_{l=1}\right],
\ee
and
\be
\label{eqn:g}
\delta g=\delta P(l-\gamma,\gamma=0)_{l=1}/\left[1-P_0(l-\gamma,\gamma=0)_{l=1}\right].
\ee

Eqn.~\ref{eqn:linearised1} constitutes a linearised equation of motion
governing the growth of the strain heterogeneity $\delta\gamma$ as a
function of strain $\gamma$ applied to the sample as a whole since the
inception of shear. Its solution is:
\be
\delta\gamma(\gamma)=\int_0^\gamma d\gamma' \delta g(\gamma')\exp\left[\int_{\gamma'}^\gamma d\gamma'' f(\gamma'')\right].
\ee
Inserting into this the forms of $f$ and $g$ from Eqns.~\ref{eqn:f}
and~\ref{eqn:g} gives, after some manipulation:
\be
\label{eqn:nearly}
\delta\gamma(\gamma)=\frac{1}{1-P_0(l-\gamma,\gamma=0)_{l=1}}\int_0^\gamma \delta P(l-\gamma',\gamma=0)_{l=1}d\gamma'.
\ee

The integral on the right hand side of Eqn.~\ref{eqn:nearly}
represents the seeding of strain heterogeneities from the fact that
the distribution of local strains, $P$, as initialised before shear
commences and subsequently advected along by shear, is not quite the
same across all streamlines, due to the finite number of elements $N$
on each streamline. It is this phenomenon that is represented by the
third term on the RHS of Eqn.~\ref{eqn:linearised}. It arises from the
disorder inherent to the amorphous material, as encoded here in the
initial distribution of local strains: any streamline that happens to
be initialised with a distribution of strains with a slightly higher
average than that across all streamlines will suffer more plastic
yielding and so strain more than the other streamlines as the level of
shear applied externally to the sample as a whole increases.

To investigate this heterogeneous seeding further, we
rewrite this  integral  in Eqn.~\ref{eqn:nearly} as
\be
\label{eqn:cumul}
\delta c(\gamma)=\delta \int_{l=1-\gamma}^1 \,dl\,P(l,\gamma=0).
\ee
We recognise this to be the standard deviation across streamlines of
the cumulative fraction of elements with initial local strains between
$l=1-\gamma$ and $l=1$, {\it i.e.}, of elements that yield within the
first $\gamma$ strain units. 

To compute this standard deviation, we note that among the $N$
elements on any streamline, a fraction
\be
c(\gamma)=\int_{l=1-\gamma}^1 \,dl\,P(l,\gamma=0).
\ee
will have strains initially in that interval, and a fraction $1-c$
will have strains initially outwith that interval. The probability of
$n$ elements having a strain in that interval is therefore given by
the binomial distribution:
\be
B(n;c,N)=\frac{N!}{n!(N-n)!}c^n(1-c)^{N-n}, 
\ee
which has normalised standard deviation:
\be
\delta c=\frac{1}{N^{1/2}}c^{1/2}(1-c)^{1/2}.
\ee
Finally, recognising that the cumulative fraction $c$ of elements with
initial strain in the interval $1-\gamma$ and $1$ will be the fraction
that yield in the first $\gamma$ strain units, we can relate $c$ to
the macroscopic stress and strain variables, $\Sigma_0$ and $\gamma_0$
via $c=\gamma_0-\Sigma_0$. This follows from the fact that the stress
$\Sigma_0(\gamma_0)$ after $\gamma_0$ strain units will be lower than
the strain $\gamma_0$ by an amount equal to the fraction of elements
that have plastically yielded. This can indeed be proved by a
straightforward integration of Eqn.~\ref{eqn:onStrain}.

The pre-factor to the integral in Eqn.~\ref{eqn:nearly} reflects that
the innately seeded heterogeneity just discussed will further be
amplified via the rebalancing of force in each strain step. Indeed,
any streamline that suffers a little more plastic yielding will strain
forward a little more in order to recover force balance, and so be
susceptible to even further plastic yielding.
Comparing this ``amplification'' pre-factor of Eqn.~\ref{eqn:nearly}
with Eqn.~\ref{eqn:stressEvolution2}, we see that it is simply equal
to the inverse slope of the loading curve of stress as a function of
strain, $1/(d\Sigma_0/d\gamma_0)$. 

Putting the above  results together gives finally:
\begin{equation}
\label{eqn:preFailure}
\delta\gamma(\gammabar)=\frac{1}{d\Sigma/d\gamma}\frac{1}{N^{1/2}}(\gamma-\Sigma)^{1/2}(1-\gamma+\Sigma)^{1/2}.
\end{equation}
We have now dropped the subscript $0$, because the stress and
strain in the homogeneous base state equal the
macroscopically measured ones, to excellent approximation, in the  early time regime in
which the strain heterogeneity remains small and our linear calculation is valid.

We pause to emphasise the significance of this result. It predicts how
strain heterogeneity develops within the sample as a function of the
strain applied on average to the sample as a whole, in the early
``pre-failure'' phase of the yielding process. The factor $1/(d\Sigma/d\gamma)$
represents an amplification factor, due to the
imposition of force balance at each strain step. This amplifies the innate statistical variation across streamlines in the initial
seeding of elemental strains, due to the disorder inherent to the
initial condition, which is described by the remaining terms. Importantly, Eqn.~\ref{eqn:preFailure} has been recast in terms of the macroscopic
global strain and stress, $\gamma$ and $\Sigma$. In this way, the degree of strain heterogeneity {\em within} the sample is related to the {\em externally} measured rheological signals.

\begin{figure}[!t]
\includegraphics[width=1.0\columnwidth]{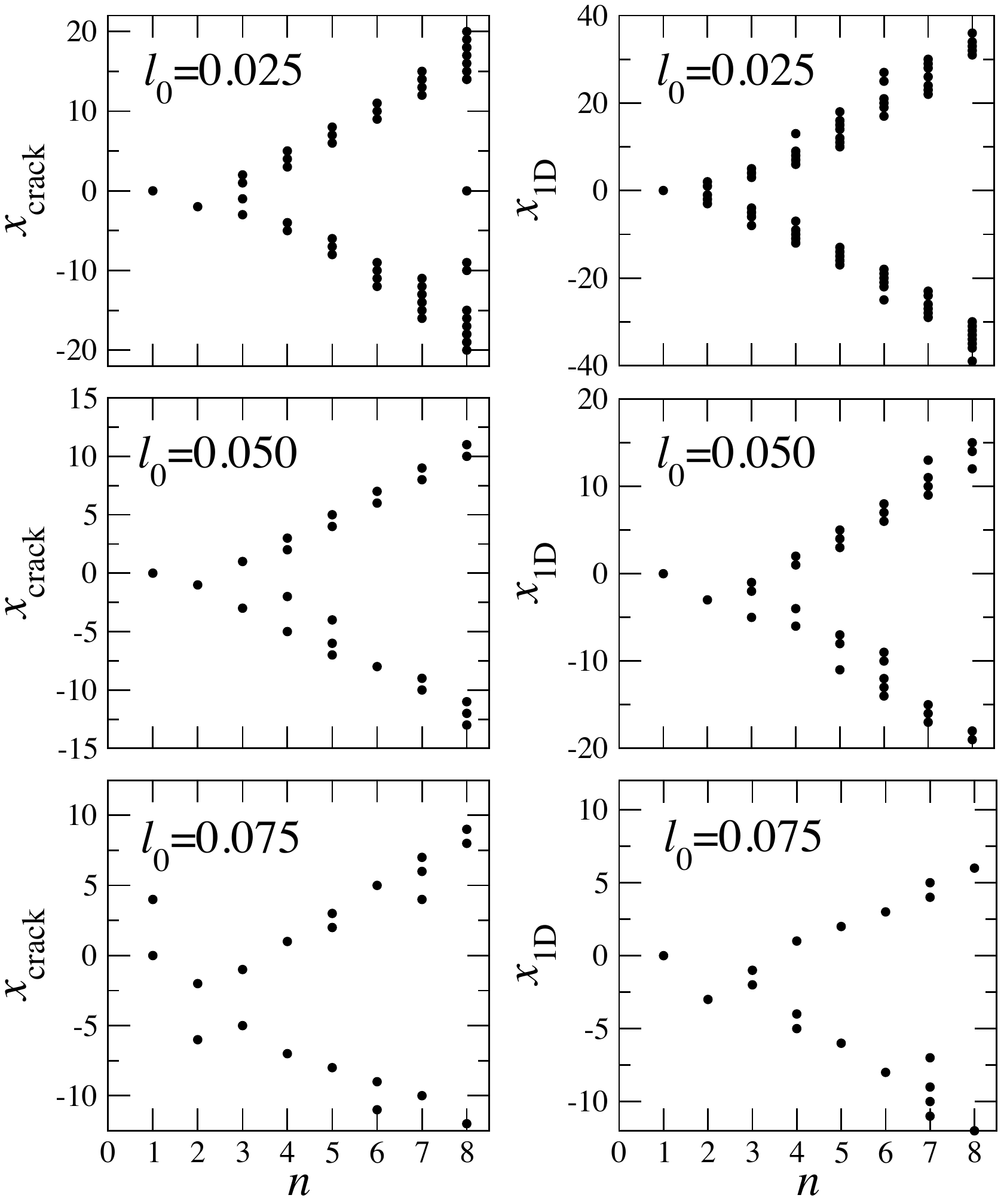}
\caption{Initial development of a shear band for values of the annealing parameter $l_0=0.025, 0.050, 0.075$ in panels downwards, and a system size $N=512$. {\bf Left)} In the full 2D elastoplastic model, with $x_{\rm band}$ denoting the coordinate ($x$ or $y$) along which the band spreads, shifted such that the initial plastic event is at the origin. {\bf Right)} In our simplified 1D elastoplastic model, also with $N=512$. As can be seen, in each case the band devlopes via two ``fronts'' that spread out in opposite directions along $x_{\rm band}$ (or its counterpart $x_{\rm 1D}$).}
\label{fig:crack}
\end{figure}
\begin{figure}[!t]
\includegraphics[width=1.0\columnwidth]{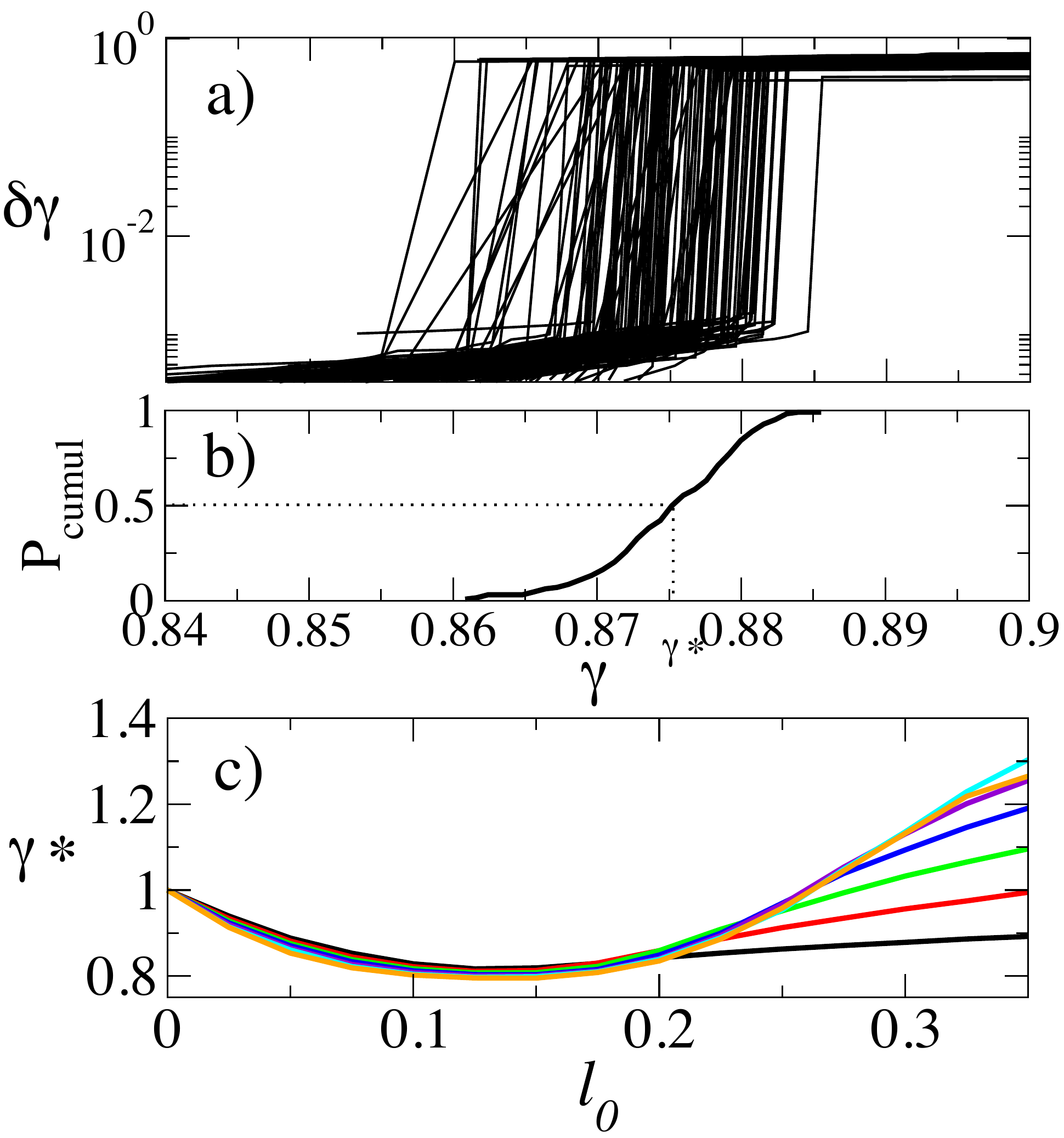}
\caption{{\bf a)} Growth of the standard deviation of strain across the sample $\delta \gamma$ as a function of overall imposed strain $\gamma$ in the prefailure then failure regime. Failure occurs at the sharp sudden upturn. Simulations are performed with a lattice size $N\times N = 256\times 256$ and an initial level of sample annealing 
$l_0=0.050$, for 128 different values of the random number seed. {\bf b)} Cumulative probability distribution of failure strains, extracted from data in a). Failure strain for each data set defined as strain at which $\delta\gamma=0.1$. {\bf c)} Average failure strain $\gamma^*$ as a function of the  annealing parameter $l_0$ (recall that small $l_0$ corresponds to strong annealing), for system sizes $N=64,128,256,512,1024,2048,4096$ in curves black, red, green, blue, violet, cyan, orange upwards at the right. $\gamma^*$ defined as strain at which $P_{\rm cumul}=0.5$.}
\label{fig:failure}
\end{figure}

In the absence of force balance -- {\it i.e.}, in the mean-field
elastoplastic model - the amplification factor described above will be
missing, but the disorder innate to the initial condition remains. In
mean field, therefore, the strain heterogeneity is predicted to grow
instead simply as
\begin{equation}
\label{eqn:preFailureMF}
\delta\gamma(\gammabar)=\frac{1}{N^{1/2}}(\gamma-\Sigma)^{1/2}(1-\gamma+\Sigma)^{1/2}.
\end{equation}
This is plotted by red dashed lines in Fig.~\ref{fig:preFailure}a) and
shows excellent agreement with our simulation results (black solid
lines) for the growth of strain heterogeneity as a function of overall
applied strain in the mean field model.

Our prediction of Eqn.~\ref{eqn:preFailure} for the 2D lattice
elastoplastic model with force balance is plotted by red dashed lines
in Fig.~\ref{fig:preFailure}b). The corresponding simulation data is
shown by black solid lines. Excellent agreement is again found between
the two early in the yielding process. It is worth noting, however,
that in this regime the difference between the mean field and full
models is anyway relatively small, because the spatial correlations
arising via force balance have not yet had chance to build. At larger
strains, once the heterogeneity is better developed, the analytical
prediction and simulation results quantitatively differ from each
other. This is to be expected, for two reasons. First, once the
heterogeneity becomes appreciable the linear assumption of the above
analysis breaks down. Second, once significant numbers of plastic
events occur, the base state will differ from the simple one assumed
in our calculation, which simply takes the elastically shifted
counterpart of the initial distribution. Nonetheless, the
amplification predicted by the factor $1/(d\Sigma/d\gamma)$ arising
from force balance is indeed present in the simulation data for the 2D
lattice model (note the upturning black lines in the top right of
Fig.~\ref{fig:preFailure}b), and absent in mean field (note the nearly
flatlining black lines in the top right of
Fig.~\ref{fig:preFailure}a).

For very poorly annealed samples, for which there is no stress
overshoot as a function of strain, the amplification factor
$1/(d\Sigma/d\gamma)$ pertaining to this pre-failure regime still
predicts a divergence in strain heterogeneity, at the level of this
linear calculation, as the stress $\Sigma$ approaches as constant at
large strains $\gamma\to\infty$, such that $d\Sigma/d\gamma\to
0$. Although nonlinear effects that become relevant once the amplitude
of the strain heterogeneity becomes significant will then be expected
to regularise any divergence, the (initially) leftmost curve of
Fig.~\ref{fig:preFailure}b) indeed confirms that appreciable strain
heterogeneity arises even for these very poorly annealed samples.

We recall finally that the $1/\sqrt{N}$ scaling in Eqns.~\ref{eqn:preFailure} and~\ref{eqn:preFailureMF} stems from the fact
that the distribution of local strains is not quite the same across all streamlines, due
to the finite number of elements $N$ on each streamline. In the limit $N\to\infty$, this difference is predicted to approach zero, giving zero strain heterogeneity in this prefailure regime (before catastrophic failure then takes over, as explored below). This is a direct consequence of the assumption made almost ubiquitously in studies of elastoplastic models: that the origin of any initial heterogeneity indeed lies in the disorder inherent to the material. In deformation geometries with a heterogeneous stress field, as pertains for example to a soft material sheared in a curved rheometer, strain heterogeneity would additionally arise from that stress heterogeneity, and would be expected to exceed that arising from the disorder in the large $N$ limit. Our calculation has assumed a regime in which heterogeneity arising from disorder exceeds that arising from systematically varying stress fields.

\subsection{Catastrophic brittle failure stage}
\label{sec:failure}

\subsubsection{2D lattice elastoplastic model}
\label{sec:fail2D}

Now we turn to a discussion of the catastrophic failure event, in which a shear band spreads quickly across the sample. Our particular aims are to predict and understand its onset in highly annealed samples

The left panels of Fig.~\ref{fig:crack} show in more detail the
initial development of the shear band along each of the top three rows in
Fig.~\ref{fig:crackMaps}, for values of the annealing parameter
$l_0=0.025,0.050, 0.075$ respectively, corresponding to highly
annealed samples. Each of these left panels in Fig.~\ref{fig:crack}
shows the coordinate along the banding direction, $x_{\rm band}$, at
which plastic events occur at successive iterations $n$ of the model
dynamics after the final strain step before banding occurs.  (Recall
that at each iteration we reimpose force balance, then yield any
elements that are taken above yield in consequence.) The coordinate
$x_{\rm band}=y$ for $l=0.025$ and $0.075$ and $x_{\rm band}=x$ for
$l=0.050$ in Fig.~\ref{fig:crackMaps}.  $x_{\rm band}$
is then shifted in Fig.~\ref{fig:crack} such that the plastic event
that initiates the band lies at the origin. In the early stage of
banding in highly annealed samples, most plastic events occur at the
same value of the coordinate orthogonal to $x_{\rm band}$, with only
a small number at the immediately adjacent coordinate value (not
shown). This justifies the 1D treatment of banding that will follow.

\begin{figure}[!t]
\includegraphics[width=1.0\columnwidth]{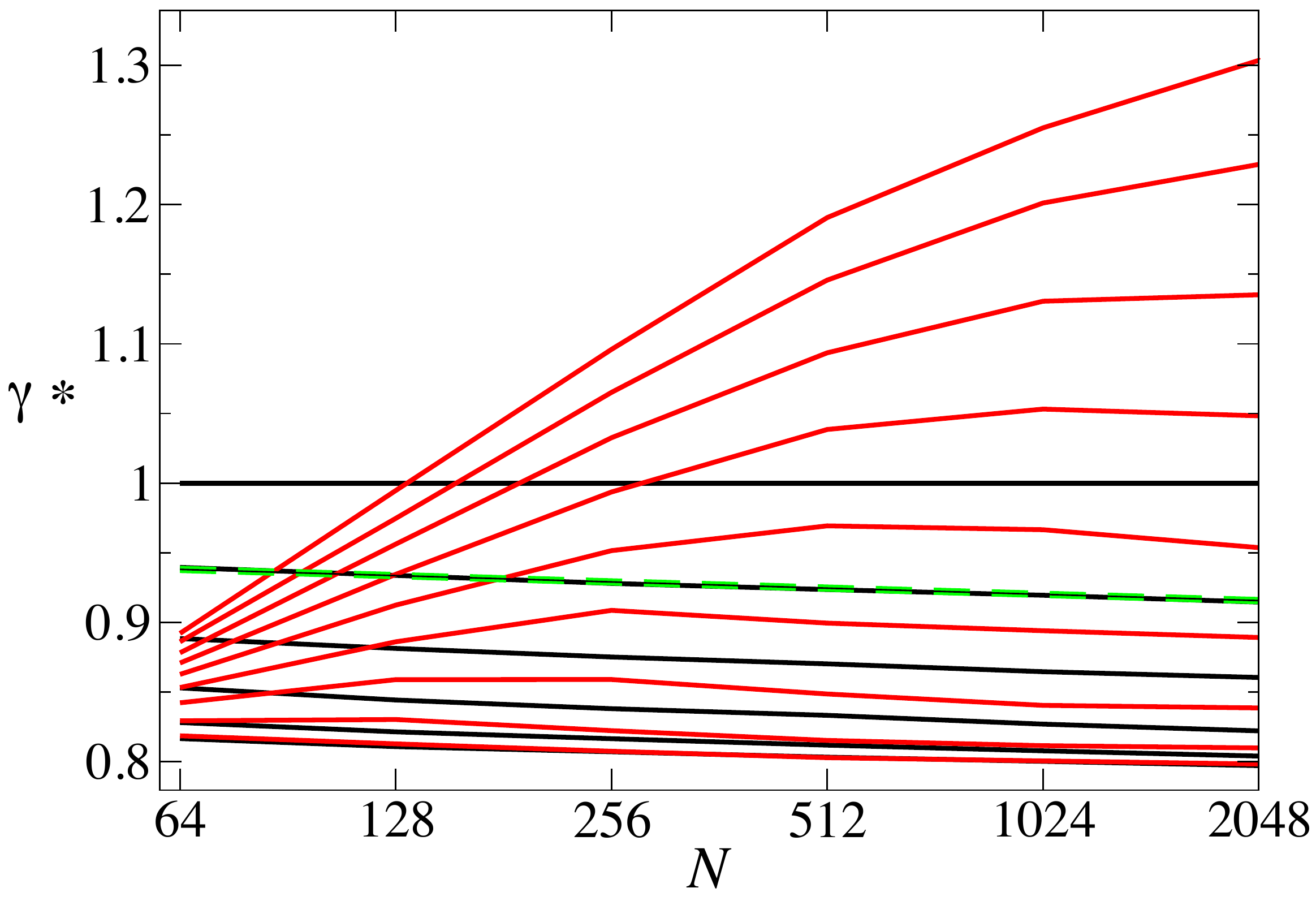}
\caption{Average failure strain as a function of the linear system size $N$ of the two-dimensional $N\times N$ lattice model for several values of the annealing parameter. Black curves  are for strong annealing, with  weaker annealing, $l_0=0.000, 0.025, 0.050, 0.075, 0.100, 0.125$, in curves downwards. Red curves are for weaker annealing, with  wearker annealing, $l_0=0.150, 0.175, 0.200, 0.225, 0.250, 0.275, 0.300, 0.325, 0.350$ in curves upwards. The green dashed line, for the strongly annealed case of $l=0.025$, is a fit to the form $\gamma^* = 0.965 - 0.00646 \log N$.}
\label{fig:size}
\end{figure}

Fig.~\ref{fig:failure}a) shows the strain heterogeneity $\delta\gamma$
across the sample as a function of the spatially averaged strain
$\gamma$, for a series of $128$ runs, each with a different random
number seed, again in a highly annealed system. In each run,
$\delta\gamma$ initially grows slowly in the ``pre-failure'' regime,
then shows a sharp sudden upturn as a band nucleates. For each of the
$128$ runs, we define the strain at which failure occurs as that at
which $\delta\gamma$ crosses the given threshold, $\delta\gamma=0.1$. We
plot the cumulative distribution $P_{\rm cumul}$ of these failure
strains in panel b). We finally extract from it the average failure
strain, $\gamma^*$, defined as the strain at which the cumulative
distribution equals $1/2$, and plot $\gamma^*$ as a function of the
level of the annealing parameter $l_0$ in c), for several different
system sizes. The same data for the average failure strain, $\gamma^*$
are plotted in Fig.~\ref{fig:size} as a function of the system size,
for different values of $l_0$.

For strongly annealed samples (small $l_0$), the failure strain
decreases (noticeably) with decreasing levels of sample annealing
(increasing $l_0$), and decreases (very weakly) with increasing sample
size $N$. This can be understood as follows. For these highly annealed
samples, the initial distribution of local strains is narrow. As the
strain applied to the sample as a whole increases, this distribution
shifts upwards. Eventually, the single element at its outmost
rightward reach (largest $l$) fails. Once this happens, all other
elements are by now near enough the yielding threshold themselves that
the effect of the first element yielding will be to trigger a
percolating cascade of further yielding events, resulting in a
shear band. Therefore, the failure strain in this regime of strong
annealing will simply be that at which the first element yields. This
in turn is equal to one (the yielding threshold itself) minus the
largest local strain initially present in the sample. This largest
initial strain will increase with decreasing initial annealing,
because the initial distribution is broader, and increase with
increasing system size, because there are more elements in the sample
to populate the extreme wings of the distribution. 
The arguments
presented in words in this paragraph are encoded in the term
$P(l-\gamma,\gamma=0)_{l=1}d\gamma$ in the calculation that follows in
Sec.~\ref{sec:failCalc} below.

For weakly annealed samples, the failure strain increases with
decreasing levels of sample annealing (increasing $l_0$) and increases
with increasing sample size $N$. This can be understood as
follows. For these poorly annealed samples, the the initial
distribution of local strains is broad. As this distribution shifts
upwards with increasing applied strain, elements start to yield. As
the first few do so, however, an insufficient fraction of other
elements are themselves close enough to threshold to provide a
percolating onwards path along which further events can be
triggered. Indeed, because of the breadth of the distribution in this
regime, a large stress needs to accumulate in the sample as a whole
before there is a percolating path for a shear band through the
system. This becomes more pronounced the larger the distribution
breadth (larger $l_0$) and the further the shear band has to propagate
(larger $N$). 

\subsubsection{Simplified 1D elastoplastic model}
\label{sec:fail1D}

Our aims now
are to understand the mechanism of shear band propagation in
Figs.~\ref{fig:crackMaps} and~\ref{fig:crack}, to predict the
probability distribution of applied strains $\gamma$ at which banding
sets in, as reported in Fig.~\ref{fig:failure}b), and thereby finally
to predict the average failure strain $\gamma^*$ in
Fig.~\ref{fig:failure}c). To this end, we shall first simulate a
simplified elastoplastic model comprising a single streamline of
elements arranged along the direction along which the band spreads.
For strong annealing, we will find this simplified 1D simulation to
predict the banding dynamics of the 2D model actually rather well. We
shall then perform a simple 1D analytical calculation, inspired by
this 1D simulation, aimed at capturing the observed behaviour.

Our simplified 1D elastoplastic model comprises $N$ elastoplastic
elements arranged along a single line in the direction $x_{\rm 1D}$
along which shear banding occurs. ($x_{\rm 1D}$ is therefore the
counterpart of $x_{\rm band}$ in the 2D model.) The dynamics of the
1D model are chosen closely to mirror those of the 2D model, as
follows. Prior to shear, each element is initialised with a local
strain drawn from the same distribution as pertained immediately prior
to shear in the 2D model. We then apply shear as follows. At each
strain step, we first enquire which is the least stable element, and
how much strain must be added to take it just over threshold. We then
add that much strain to all elements, and to the global strain
variable. That least stable element is now just above threshold, and
is yielded. All elements then have their strain adjusted according to
a 1D stress propagator that we discuss in the next paragraph, centred
on the element that has just yielded. This process of stress
propagation may then take some other elements above threshold. Those
elements are then yielded, and the strain of all elements again
adjusted via a superposition of our 1D stress propagators, each
centred on one of the newly yielded elements. This process is repeated
iteratively until no elements are left above threshold. We then
proceed to the next strain step.

\begin{figure}[!t]
\includegraphics[width=1.0\columnwidth]{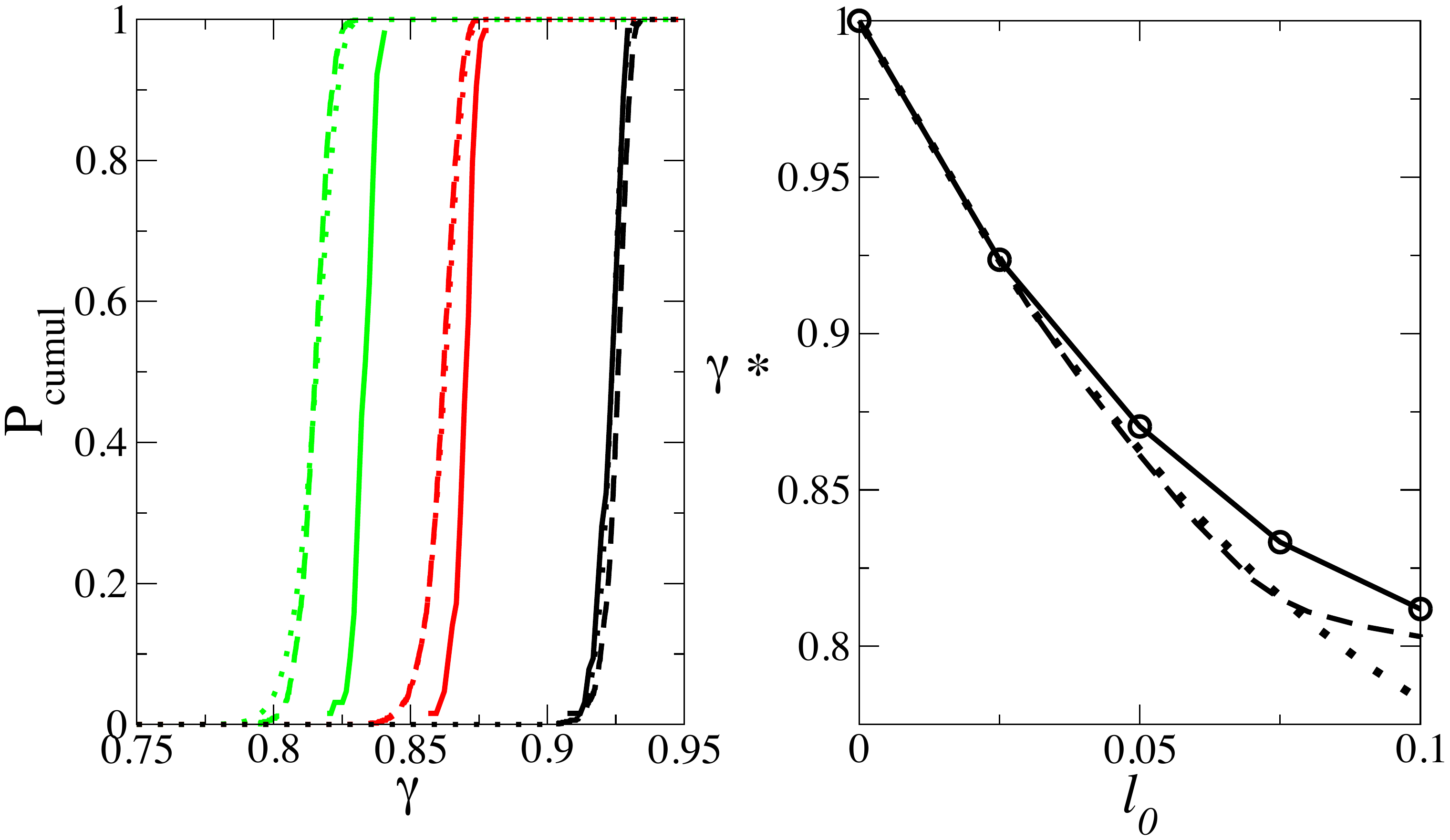}
\caption{{\bf Left)}  Solid lines: cumulative probability distribution of failure strains in the full 2D simulation with a lattice of size $N\times N=512\times 512$, calculated over many runs, each with a  different random numbers seed. Dashed lines: corresponding distributions for the simplified 1D model with $N=512$. Dotted lines:  distribution predicted using Eqns.~\ref{eqn:cumulDistr} and~\ref{eqn:probRunaway}. Anealing parameter $l_0=0.025$ (black), $l_0=0.050$ (red) and $l_0=0.075$ (green)  in curve groups leftwards.
 {\bf Right)} Corresponding average failure strain plotted as a
 function of $l_0$, for the 2D model (solid line), 1D model (dashed
 line), and predicted using Eqns.~\ref{eqn:cumulDistr}
 and~\ref{eqn:probRunaway} (dotted line). Recall that strongly annealed
 samples correspond to small $l_0$. }
\label{fig:failStrain}
\end{figure}

The 1D propagator to which we referred in the previous paragraph is
calculated from the full Eshelby propagator of the 2D lattice by
taking the values of this function along one dimension only, at the
origin of the other dimension. Therefore, if the stress propagator
following an event at $x=x_0,y=y_0$ on the 2D lattice is
$\tilde{E}(x-x_0,y-y_0)$, the stress propagator following an event at
$y=y_0$ on the 1D lattice is $\tilde{E}(0,y-y_0)\equiv E(y-y_0)$. We
observe numerically that the function $E(y)$ is well fit to the form
$0.309y^{-2}$ for $y>2$ and $y
\ll N$, up to smaller variations on the spacing of the lattice
site. For values of $y$ approaching $N$, $E$ departs from the power
law decay to obey the periodic boundary conditions. $E(1)=0.079$.

In the regime of strong annealing (small $l_0$), each run of this 1D
model shows a failure event that closely resembles that seen in the
full 2D simulation with matched $l_0$, at least qualitatively. This can
be seen by comparing each left panel of Fig.~\ref{fig:crack} for the
2D model with its counterpart in the right panel for the 1D model. For
any $l_0$, we run this 1D code $N$ times, each with a different random
number seed, to account for the fact that there are $N$ streamlines in
the 2D model. The minimum failure strain across these $N$ separate
runs is then taken as the prediction of the 1D model for the strain at
which failure would occur in the 2D model. Running that $N$-fold
simulation $M$ times, we construct finally the cumulative probability
distribution of failure strains $P_{\rm cumul}(\gamma)$, across those
$M$ runs. From this we calculate finally the average failure strain
$\gamma^*$.

The cumulative distribution of failure strains is shown for three
different levels of annealing in Fig.~\ref{fig:failStrain}a). Solid
lines show results for the 2D model, and dashed lines for the 1D
model. For the most highly annealed samples, the correspondence
between the 1D and 2D models is excellent.  For less well annealed
samples, we find progressively less good agreement. The average
failure strain is shown as a function of the annealing parameter $l_0$
in Fig.~\ref{fig:failStrain}b), by the solid and dashed line for the
2D and 1D models respectively. As can be seen, the average failure
strain of the 1D model agrees reasonably with that for the 2D model
even for samples as poorly annealed as $l_0\approx 0.1$, with full
agreement in the strongly annealed limit, $l_0\to 0$.

\subsubsection{Simplified 1D analytical calculation}
\label{sec:failCalc}

We seek finally to understand the onset of shear banding via an analytical
calculation that is inspired by our simplified 1D elastoplastic model,
as follows.

In any strain increment $\gamma\to \gamma+d\gamma$ since the inception
of shear, the probability of any given element locally yielding is
$P(l,\gamma)_{l=1}d\gamma$. In the early stages of shear, recall that
$P(l,\gamma)=P(l-\gamma,\gamma=0)$ to good approximation, because the
initial distribution that pertained before shearing commenced has
simply been shifted rightward by elastic loading, with only small
changes due to plastic relaxation, which are neglected here. In the
limit of small strain increment $d\gamma\to 0$, the probability of
exactly one out of $N^2$ elements yielding, $P_1(\gamma)$, is
\be
N^2 P(l-\gamma,\gamma=0)_{l=1}d\gamma.
\ee
(The $N^2$ elements can be viewed either as the elements in $N$ runs
of our 1D code with $N$ elements along a line; or as the $N\times N$
elements on the counterpart 2D lattice. The probability of two
elements yielding is $O(d\gamma^2)$ and can be neglected in the limit
$d\gamma\to 0$.)

Let us now denote by $r(\gamma)$ the probability that, when an
element yields in the strain interval $\gamma\to\gamma+d\gamma$, it leads to
a runaway shear band and therefore sample failure. The probability that
a shear band arises in $\gamma\to\gamma+d\gamma$ is then
$N^2r(\gamma)P(l-\gamma,\gamma=0)_{l=1}$. Denoting by $I$ the
probability that no shear band has  arisen by a strain of $\gamma$, we
have
\be
\label{eqn:decay}
\frac{dI}{d\gamma}=-N^2r(\gamma)P(l-\gamma,\gamma=0)_{l=1}I.
\ee
This can be integrated to find $I$. The probability $1-I$ that a shear band
will have arisen by a strain $\gamma$, {\it i.e.}, the cumulative
probability distribution of failure strains, is then:
\be
\label{eqn:cumulDistr}
P_{\rm cumul}(\gamma)=1-\exp\left[-N^2 \int_0^\gamma d\gamma'r(\gamma')P(l-\gamma',\gamma=0)_{l=1}\right].
\ee

It remains finally for us to consider the probability $r(\gamma)$
that, when a single element yields in the interval
$\gamma\to\gamma+d\gamma$, it leads to the development of a runaway
shear band and therefore to sample failure. 

Imagine a shear band that has already grown to comprise $n$ contiguous
events, centred on the plastic event that triggered it initially. What
is the probability of an event occurring subsequently a distance $m$
lattice sites away, on one side of this band? Such a site would, at
the time the band first nucleated, have had a strain $\gamma$. It now
has a strain
\be
\label{eqn:contiguous}
\gamma'=\gamma+\sum_{n'=1}^nE(m+n'),
\ee
due to the effects of stress propagation from each of the $n$ sites
that have already yielded along the band. Here $E$ is the stress
propagator used in the 1D elastoplastic model, as described
previously. The probability of this site now yielding, $p_{nm}$, is
the probability that its initial strain before shearing commenced had
been in the interval $1-\gamma'$ to $1-\gamma$, which can be written as
\be
\label{eqn:macro2}
p_{nm}=c(\gamma')-c(\gamma).
\ee
Here we use the same definition of the cumulative distribution as in
Eqn.~\ref{eqn:cumul}.
The probability of zero elements on either side
of the band of length $n$ then yielding is 
\be
\sum_{m=1}^M(1-p_{nm})^2,
\ee
in which the square accounts for the fact that there are two sides,
one to the left of the pre-existing band, and one to the right. The
probability that at least one event occurs and the band keeps
spreading  is then
\be
1-\sum_{m=1}^M(1-p_{nm})^2.
\ee
The probability of the band developing to a large number of events
$N_{\rm events}$, giving a runaway instability, is then
\be
\label{eqn:probRunaway}
r(\gamma)=\prod_{n=0}^{N_{\rm events}}\left[1-\sum_{m=1}^M(1-p_{nm})^2\right].
\ee

Inserting this expression~\ref{eqn:probRunaway} for the probability
that a single plastic event occurring at strain $\gamma$ triggers a
runaway banding instability into Eqn.~\ref{eqn:cumulDistr}, and
calculating this integral numerically, we find the cumulative
distribution of failure strains given by the dotted lines in
Fig.~\ref{fig:failStrain}, left). (We have taken the limit of large
$N_{\rm events}$ and $M$, but in fact already found good convergence to this limit
by $N_{\rm events}=16$ and $M=2$.) As can be seen, it shows excellent agreement
with the the cumulative distribution of failure strains extracted from
the simplified 1D elastoplastic model, which in turn shows good
agreement with the corresponding distribution for the full 2D
elastoplastic model across the range of $l_0$ values shown, with
excellent agreement for the smallest $l_0$. The correspondingly
predicted averaged failure strain $\gamma^*$ is plotted as a function
of the annealing parameter $l_0$ by the dotted line in
Fig.~\ref{fig:failStrain}, right), along with that from the 1D model
simulation (dashed line) and the 2D model simulation (solid line).

While we believe the derivation up to Eqn.~\ref{eqn:cumulDistr} to be
rigorous, up to corrections that we have discussed, the subsequent
arguments used to obtain $r(\gamma)$ contain more severe
assumptions. For example, Eqn.~\ref{eqn:contiguous} assumes a
proto-band of fully contiguous plastic events along a line,
whereas in practice, successive events can skip over some lattice
sites to leave them unyielded.  (Recall Fig.~\ref{fig:crack}.) 
Nonetheless, as discussed in the previous paragraph, this simple
calculation gives convincing agreement with the results of our 1D
simulations, at least for values of the annealing parameter up to
$l_0=0.075$, and which in turn agree well with our full 2D simulations for
the most strongly annealed samples.

Although the expressions ~\ref{eqn:cumulDistr}
and~\ref{eqn:probRunaway} that determine the probability distribution
of failure strains look rather complicated, the quantities upon which
they depend, $p_{nm}$ and $P(l-\gamma,\gamma=0)_{l=1}$, are together
determined completely by the parameter $l_0$ that describes the level of disorder in the sample, as determined by the degree of sample annealing prior to shear, and  the system size, $N$.

It is worth pausing to emphasise the significance of this result,
which applies for materials described by the elastoplastic model
studied here, in the limit of strong annealing and in quasistatic
shear. It predicts the probability distribution of applied strain
values at which catastrophic sample failure occurs, in terms of the
disorder inherent in the sample prior to shear, as determined by the
degree of sample annealing, and the system size $N$. It therefore
indicates a possible way to predict catastrophic material failure
before it occurs.

\section{Discussion: implications for experiment and molecular simulations}
\label{sec:discussion}

In this section, we discuss the implications of our findings for experiment and molecular simulation.

Amorphous materials can be broadly subdivided into two (limiting, idealised) subcategories:
``athermal'' and ``thermal''.  Athermal materials comprise
substructures (foam bubbles, emulsion droplets, etc) large enough that
Brownian effects can be discarded, with energy barriers
impeding bubble/droplet rearrangement that greatly exceed $k_{\rm B}T$. Commonly studied experimental examples include dry
granular packings, dense granular suspensions, foams, and emulsions. In thermal materials, on the other
hand, the constituent substructures and/or local energy barriers are
small enough that thermal fluctuations play an important
role. Colloidal and polymeric glasses, colloidal gels, molecular and
metallic glasses typically fall into this second class, although deep within the glass phase, far below the glass transition temperature, energy barriers will  be large on the scale of $k_{\rm B}T$. The athermal model studied here is accordingly expected to apply primarily to materials with substructures large enough that Brownian motion can be neglected upfront, as well as molecular and metallic glasses deep within their glass phase.

Indeed, at any finite temperature $T$, the timescale for spontaneous particle rearrangement in the absence of shear, $\tau_{\rm thermal}$, is expected to scale as $\tau_0\exp(E/k_{\rm B}T)$, with $E$ the typical energy barrier to rearrangement, and $\tau_0$ the microscopic timescale for local intracage particle motion. In shear, groups of particles are expected to reach the threshold for local yielding on a timescale $\tau_{\rm shear}$ that scales as $\sqrt{E/k}/\dot{\gamma}$, {\it i.e.,} as the time required to strain of order $\sqrt{E/k}$ at a strain rate $\dot{\gamma}$. (Recall that $k$ is the local modulus.) Once the threshold is reached, yielding is then expected to take place on the microscopic timescale $\tau_0$. The protocol of athermal quasistatic shear studied here pertains to the regime of well separated rates $\tau_{\rm thermal}^{-1}\ll \tau_{\rm shear}^{-1}\ll \tau_0^{-1}$. In fully thermal materials, treating $\tau_{\rm thermal}^{-1}$ and  $\tau_{\rm shear}^{-1}$ as well separated will not be valid. Indeed, as one moves way from the athermal limit and towards the thermal regime, activated processes will play an increasing role, avalanches associated with yielding are likely to start to overlap, and the yielding transition may be somewhat smoothed compared with that explored here. This would be an interesting topic for future study.

In the context of hard materials, the trend towards increasing fracture toughness with decreasing system size that we find for moderately to well annealed samples is consistent with experiments on nanoglass and metallic glasses~\cite{wang2016sample}, where it was suggested
to arise from the reduced number of sites for the nucleation of shear
bands as system size decreases. Increasing ductility with decreasing sample size was also reported experimentally in Ref.~\cite{jang2010transition}, with smaller samples being argued to require a larger stress for the nucleation of shear bands.  A trend towards increasing ductility with decreasing sample size in samples subject to compressive loading was also discussed in ~\cite{yang2012size}, along with an important difference between failure in compressive and tensile loading, which remains an open challenge for future theoretical work. Decreasing fracture toughness with decreasing ``fictive temperature" (increasing annealing)~\cite{rycroft2012fracture} has recently been reported in metallic glasses~\cite{ketkaew2018mechanical}.  

In the context of soft materials, our theoretical predictions are consistent with and indeed potentially explain the widespread experimental observation that these materials typically yield in a less abrupt way than hard materials. A key finding of our work has been that abrupt yielding tends to occur, for modestly annealed samples, in the limit of infinite system size (compared with the typical size of a material's constituent molecular or mesoscopic particles) and slow shear rate (compared with the inverse of the intrinsic material timescale). In comparison with hard materials, soft materials typically have much larger constituent mesostructures: the droplets in an emulsion are much larger than the constituent particles of a metallic or molecular glass glass, for example. Accordingly, their associated timescales are typically much slower. The crossover to the abrupt yielding that we predict for large system sizes and slow shear rates is accordingly likely to be accessed only at larger system sizes and slower shear rates for soft materials than hard materials, quite possibly outside the window explored in many existing experimental works.

Nonetheless, a growing body of experimental data increasingly shows that soft materials can also show brittle failure~\cite{ISI:000381495100025,ISI:A1996TR32200033,ISI:000073082400044,ISI:000334096300032,ISI:000265285700042,ISI:000341260100009}. Such observations are emerging rapidly in the light of new experimental techniques that can now directly access the microscopic precursors to
sudden failure~\cite{ISI:000429012500051,ISI:000392096800036,ISI:000341025700007}. Our results predicting the way a microscopic precursor -- an initially localised plastic event -- leads to a cooperatively growing shear band are particularly pertinent in the context of these new techniques. This connection with be explored in further detail in future work.

We have also made detailed quantitative predictions for the slow initial growth of strain heterogeneity within a material after shearing first commences, prior to any regime of final catastrophic  material failure. Of particular relevance to experiment is our elucidation of the way in which this strain heterogeneity {\em within} a sample is predicted via the macroscopically measurable {\em external} rheological signals of strain and stress.
This prediction is directly testable in flow velocimetry experiments of the kind that have been pioneered in studies of shear banding in soft materials such as emulsions and gels in recent years~\cite{manneville2008recent}. Specifically, such experiments are now capable of measuring the degree of strain rate heterogeneity across the sample as a function of time, and time-integrating this signal to obtain the growing degree of strain heterogeneity. This can then be directly compared with our prediction in Fig.~\ref{fig:preFailure}b and Eqn.~\ref{eqn:preFailure} above by performing in-tandem measurements of the macroscopic stress-strain curve.

Our findings also have important implications for the interpretation of molecular simulations (in silico experiments) of low temperature glasses, for example as reported in Ref.~\cite{singh2020brittle}. In particular, the data in Fig. 3a of that work are consistent with the scenario reported here for moderately or poorly annealed samples, with ductile yielding for small system sizes and a trend towards brittle yielding for larger system sizes. Our work predicts that the same simulation curves reported for even more strongly annealed samples should show a crossover towards brittle yielding at smaller system sizes. This predition is directly testable, and is indeed consistent with some very recent molecular simulation results~\cite{david2021finite}. 

We have already made detailed discussion of our work in relation to the molecular simulations of Ref.~\cite{ozawa2018random} in Sec.~\ref{sec:brittle} above, and we shall not repeat this here.

\section{Conclusions}
\label{sec:conclusions}

In summary, we have studied theoretically the yielding of slowly
sheared athermal amorphous materials. Our principal
contributions have been to predict the conditions under which yielding
will be ductile or brittle; to understand the initial growth of strain
heterogeneity that is a precursor to material failure; to elucidate
the way in which the nucleation and propagation of a shear band leads
finally to catastrophic failure; and to predict the distribution of
applied shear strains at which catastrophic failure will occur, in
terms of the disorder inherent in the sample, as determined by the
degree of sample annealing, and the system size. 

For highly annealed samples, we have found yielding to be brittle for
all samples sizes. In contrast, poorly annealed samples show an
important dependence on the size of the sample of material being
sheared, with apparently ductile yielding for small samples, and
brittle yielding for large samples.  This has important  implications for experiment, 
in predicting a tendency towards increasing brittleness (for a fixed sample size) with increasing annealing; and that materials subject to a given level of annealing will show a different mode of failure, depending on the size of the sample being deformed. 

Our work has shown that yielding comprises two distinct
stages: a ``pre-failure'' stage, in which small levels of strain
heterogeneity slowly accumulate within the material, followed by a
catastrophic brittle failure event, in which a shear band quickly
propagates across the sample via a cooperating line of plastic events.

In the pre-failure regime, we have provided an exact analytical expression
for the slowly growing level of strain heterogeneity, expressed in
terms of the macroscopically measurable stress-strain curve and the
sample size, and in excellent agreement with our simulation results.

We have further elucidated the mechanism of subsequent
catastrophic material failure, in which a shear band
nucleates and spreads quickly across the sample. For highly annealed
samples, our simulations have shown that that a single element first
yields plastically and, in elastically propagating its stress to other
elements via the Eshelby propagator, creates further nearby yielding
events, leading quickly in turn to a percolating chain of events along
a line~\cite{hieronymus2017shear}.

In its use of periodic boundary conditions, our
work necessarily pertains to the homogeneous nucleation of a shear band, triggered by
plastic events that arise within the body of the disordered material,
and which determines the ultimate strength of a material in the
absence of surface imperfections. Clearly, it is imperative to understand this simpler case of homogeneous nucleation in order to elucidate the ultimate toughness of a material without surface imperfections; and also as a foundation on which to build an understanding of  heterogeneous nucleation. In future, it would be interesting
to simulate samples with external borders that have imperfections and
indentations~\cite{rycroft2012fracture}, to study the heterogeneous nucleation of shear bands
arising in regions of concentrated stress. We have also
limited ourselves to shear deformations, and it would be interesting
in future to consider other deformation and loading protocols, such as
planar extension.

It remains an open challenge to discriminate between or unify the picture of yielding elucidated here and those put forward in earlier studies, for example within a replica
field
theory~\cite{ISI:000370815100008,ISI:000402296700034,ISI:000410885300004}; as a directed
percolation transition~\cite{ISI:000386386400004,ISI:000384392300003};
within a random first order transition theory for the glass
transition~\cite{ISI:000309611400031}; as a Gardner transition
~\cite{rainone2015following}; as a spinodal
point~\cite{urbani2017shear}; and within particle simulations that
seed initial weak spots~\cite{ozawa2021rare}.

{\it Acknowledgements ---} This project has received funding from the European Research Council (ERC) under the European Union's Horizon 2020 research and innovation programme (grant agreement No. 885146).


%

\end{document}